\documentclass[12pt, draftclsnofoot, journal, onecolumn]{IEEEtran}
\usepackage[T1]{fontenc}
\usepackage{aecompl}
\usepackage{blindtext}
\usepackage{caption}
\usepackage{comment}
\usepackage{lipsum}
\usepackage{amsmath}
\DeclareMathOperator*{\argmax}{arg\,max}
\usepackage{amssymb}
\usepackage{setspace}
\usepackage{bm}
\usepackage{mdwmath}
\usepackage{graphicx}
\usepackage{caption}
\usepackage{xcolor}
\usepackage{subfigure}
\usepackage{stfloats}
\usepackage{makecell}
\usepackage{multirow}
\usepackage{acronym}
\usepackage[noadjust]{cite}
\usepackage{url}
\usepackage[linesnumbered,ruled,vlined,boxed]{algorithm2e}
\usepackage{algpseudocode}
\usepackage{cases}
\usepackage{makecell}
\newcounter{MYalgorithmic}

\newtheorem{definition}{\textbf{\textit{Definition}}}

\newtheorem{proposition}{\textit{\textbf{Proposition}}}

\newcommand{\upperroman}[1]{\uppercase\expandafter{\romannumeral#1}}

\usepackage{enumerate}
\usepackage{bbm}
\usepackage{empheq}
\begin{document}
\title{Multi-Dimensional Multiple Access with Resource Utilization Cost Awareness for Individualized Service Provisioning in 6G}

\author{Jie Mei,~\IEEEmembership{Member,~IEEE},
Wudan Han,
Xianbin Wang$^{*}$,~\IEEEmembership{Fellow,~IEEE},~and\\
H. Vincent Poor,~\IEEEmembership{Fellow,~IEEE}
\vspace{-3em}
\thanks{Jie~Mei, Wudan~Han and Xianbin~Wang are with Electrical and Computer Engineering, Western University, London, ON N6A 5B9, Canada (E-mail: \begin{scriptsize}\textsf{\{jmei28, whan55, xianbin.wang\}@uwo.ca}\end{scriptsize}). \textit{Corresponding author: Dr. Xianbin Wang.}}
\thanks{H.~Vincent~Poor is with the Department of Electrical and Computer Engineering, Princeton University, Princeton, NJ 08544, USA (E-mail: \begin{scriptsize}\textsf{poor@princeton.edu}\end{scriptsize}).}
\thanks{The MATLAB code used to obtain the numerical results in Section \ref{Simulation} is available on GitHub: \begin{scriptsize}\textsf{https://github.com/wzmdaisy/Multi-Dimensional-Multiple-Access}\end{scriptsize}.}
}
\maketitle
\begin{abstract}
The increasingly diversified Quality-of-Service (QoS) requirements envisioned for future wireless networks call for more flexible and inclusive multiple access techniques for supporting emerging applications and communication scenarios. To achieve this, we propose a multi-dimensional multiple access (MDMA) protocol to meet individual User Equipment's (UE's) unique QoS demands while utilizing multi-dimensional radio resources cost-effectively. In detail, the proposed scheme consists of two new aspects, i.e., selection of a tailored multiple access mode for each UE while considering the UE-specific radio resource utilization cost; and multi-dimensional radio resource allocation among coexisting UEs under dynamic network conditions. To reduce the UE-specific resource utilization cost, the base station (BS) organizes UEs with disparate multi-domain resource constraints as UE coalition by considering each UE's specific resource availability, perceived quality, and utilization capability. Each UE within a coalition could utilize its preferred radio resources, which leads to low utilization cost while avoiding resource-sharing conflicts with remaining UEs. Furthermore, to meet UE-specific QoS requirements and varying resource conditions at the UE side, the multi-dimensional radio resource allocation among coexisting UEs is formulated as an optimization problem to maximize the summation of cost-aware utility functions of all UEs. A solution to solve this NP-hard problem with low complexity is developed using the successive convex approximation and the Lagrange dual decomposition methods. The effectiveness of our proposed scheme is validated by numerical simulation and performance comparison with state-of-the-art schemes. In particular, the simulation results demonstrate that our proposed scheme outperforms these benchmark schemes by large margins. 
\end{abstract}

\begin{IEEEkeywords}
6G, multi-dimensional multiple access, resource utilization cost, individualized QoS provisioning.
\end{IEEEkeywords}

\IEEEpeerreviewmaketitle


\section{Introduction}
\label{sec:Introduction}
\subsection{Motivation}
The continuous momentum of data traffic growth and dramatic expansion of diverse services and vertical applications are bringing many challenges to the development of the envisioned 6-th generation (6G) wireless networks. On the one hand, with the anticipated seven-fold growth of mobile data traffic in 2022 compared to 2017~\cite{Kenneth}, future 6G networks are expected to support significantly higher data rates. On the other hand, a wide variety of emerging services and applications are expected to be supported by 6G, e.g. smart manufacturing, augmented reality, which require diverse, application-specific and individualized service provisioning in terms of data rate, latency, reliability, and power consumption~\cite{6g_cases}, \cite{you2021towards}. 
\par
Given the heterogeneity and dynamic resource constraints of future networks and wireless devices, designing highly efficient and intelligent multiple access techniques becomes critical for 6G. As a result, individualized quality-of-service (QoS) provisioning, rather than current scenario-specific solutions adopted in 5G, as well as the cost-effectiveness of User Equipment (UE) operation, are envisioned as the key features of 6G to fulfill its role as a multipurpose platform and foundation of a connected society \cite{Liu2}. Current attempts for achieving individualized QoS provisioning in 6G are primarily focused on incorporating recent technological advancements in operation/management, e.g., network slicing and mobile edge computing \cite{AI-NS, Ruitao}, which may not be practical due to the increased system complexity. Meanwhile, by adopting a bottom-up alternative, we believe that the design of next generation of multiple access schemes that can efficiently utilize multi-dimensional radio resources with situational awareness could play a vital role in individualized service provisioning for 6G \cite{liu2021application,liu2021evolution}.
\par
In achieving individualized QoS provisioning cost-effectively, next generation multiple access protocols are expected to offer finer servicing granularity for diverse QoS provisioning by exploiting multi-dimensional radio resources, including frequency, time, space, power and code domains, while considering the specific resource conditions and specific QoS requirements of each UE. To this end, several additional factors, e.g., UE hardware capabilities, and radio resource utilization cost, have to be considered to improve the effectiveness of intelligent multiple access for individualized QoS provisioning. Recently, several new multiple access techniques, particularly multiple-input multiple-output non-orthogonal multiple access (MIMO-NOMA) \cite{Dai} and rate-splitting multiple access (RSMA) \cite{mao2018rate}, have been proposed to explore additional degrees of freedom in spatial and power domains to improve spectral efficiency and system multiplexing capability. However, these multiple access schemes follow the existing scenario-specific practice in 5G and still face many challenges to fulfill the individualized demands and specific situation of each UE due to the failure for addressing the following three crucial issues:
\begin{itemize}
    \item UE-specific resource constraints and heterogeneous resource utilization costs in different domains. Each UE is inherently limited by distinctive constraints and utilization costs in multiple radio resource domains. Ideally, all UEs are expected to be equipped with powerful processing capabilities and abundant resources. However, in practice, UEs have heterogeneous hardware capabilities, including signal processing/computing capability, storage limits, and power/battery supply, which results in inherent hardware constraints and heterogeneous radio resource utilization cost in different domains \cite{PD_NOMA_cost, 9152055}. For instance, some low-cost devices have poor successive interference cancellation (SIC) capability due to limited computing capabilities and power supply, which restricts their performance for employing the power-domain NOMA. 
    \item UE-specific perceived value of radio resources in different domains. Each UE experiences a different level of resource availability, constraint, and quality for  multi-dimensional radio resources due to the distinctive channel conditions sensed by the UE. The perceived value of allocated multi-dimensional radio resources will be UE-specific due to different UEs' capabilities and induced cost when utilizing such resources. This observation offers a new perspective of intelligent multiple access design to achieve more individualized, opportunistic and optimal multi-dimensional radio resources allocation among coexisting users.
    \item UE-specific, diverse, and individualized QoS requirements. Through coarsely classifying all services into Enhanced Mobile Broadband (eMBB), massive Machine Type Communication (mMTC), and ultra-Reliable and Low Latency Communication (uRLLC), the scenario-specific multiple access schemes in 5G networks face many challenges for satisfying the dramatically increased service heterogeneity and diversity due to the wide variety of applications and devices \cite{Mei1}. Consequently, new 6G designs are expected to meet the distinctively different QoS requirements from each UE. The UE-specific  QoS requirements can be translated into resource requirements in different radio resource domains. This inspires us to explore intelligent radio resource allocation in different domains for more effective multiple access and individualized service provisioning designs.
\end{itemize}
\par
Motivated by these observations, we aim to create a multi-dimensional multiple access (MDMA) scheme, which can flexibly and opportunistically orchestrate multi-dimensional radio resources by unifying orthogonal multiple access (OMA), power-domain NOMA, and spatial-domain NOMA based on the individualized communication needs of each UE. Specifically, this paper aims to achieve the following two technical goals: a) jointly exploiting distinctive situations and disparate constraints in multiple radio resource domains among coexisting users to improve overall network communication outcome; and b) enabling individualized service provisioning for each UE by comprehensively considering both the gains achieved by fulfilling UE-specific QoS demands and the utilization cost of multi-dimensional radio resources at the UE side.
\subsection{Related Works}
With ongoing wireless evolution, future 6G networks are expected to integrate more advanced techniques for supporting services beyond current mobile use scenarios. To meet this objective, researchers have focused on different ways of enhancing multiple access techniques for 6G. These efforts can be roughly classified into two directions: a) joint utilization of all dimensional radio resources in a non-orthogonal paradigm; and b) design of new multiple access scheme considering the multi-dimensional radio resource utilization costs. In the following, a concise literature review for two aspects mentioned above is presented with involved details and designing principles. We then discuss our solution that intelligently reflects all these considerations and objectives.
\par
Joint utilization of multi-dimensional radio resources in a non-orthogonal paradigm offers significant potential for improving network performance. Following this direction, researchers designed a hybrid multiple access method by jointly engaging NOMA and OMA for the purpose of improving resource utilization and access efficiency under different network conditions \cite{9217161}. MIMO-NOMA scheme combining both spatial-division multiple access (SDMA) and power-domain NOMA are adopted to reduce the inter-user interference and improve system throughput~\cite{9148204,Dai,MIMO-NOMA}. The authors in \cite{8740921} investigated a multi-dimensional resource allocation strategy in a multi-cell multi-subcarrier downlink MIMO-NOMA network, including a spatial user-clustering design and an optimization algorithm for summed utility maximization. Simulation results demonstrate MIMO-NOMA outperforms multiple access schemes that only multiplex users in a single domain. On a parallel pathway, the authors in \cite{mao2018rate} proposed a rate-splitting multiple access (RSMA) scheme, which treats SDMA and NOMA as two special cases by combining linearly precoded rate splitting at the transmitters and SIC modules at the receivers. Furthermore, the works \cite{9461768,RSMA_JSAC} studied the sum-rate maximization issue in a single cell downlink network that employs RSMA to serve multiple users simultaneously, in which the numerical results demonstrate performance improvement compared with traditional NOMA schemes. Though these works provide a generalized framework for multi-dimensional radio resource utilization, there is a lack of analysis for resource characteristics, perceived value and utilization cost in different resource domains experienced by individual users. Furthermore, it is also worth noting that overall multi-dimensional multiple access strategies have not been systematically analyzed.
\par
However, most of existing multiple access schemes assumes that users exhibit no hardware constraints for the implementation, which is unrealistic in real environments. To address this issue, a new direction is to model the utilization cost of user devices due to the hardware capabilities and channel conditions and consider it in the design of new multiple access schemes. In \cite{PD_NOMA_cost}, a novel utility function that takes account of "costs" for access technologies is proposed to reflect the complexity of power-domain NOMA imposed on different users. Furthermore, in our previous work \cite{Liu_TWC}, we propose a novel metric called total non-orthogonality to estimate the power consumption at all UE devices caused by adopting NOMA. The total non-orthogonality consists of non-orthogonal resource allocation effects in all resource domains of whole network, e.g. the spatial-domain non-orthogonality and power-domain non-orthogonality. Instead of optimizing the network-wide performance, our work \cite{Wudan} designed a user-centric multi-dimensional resource allocation scheme, which manages to maximize each UE's QoS performance and minimizes multi-domain non-orthogonal interference at the UE side. However, this work does not consider distinctively different characteristics of coexisting UEs, including their different service requirements, channel conditions, and hardware capabilities. Thus, new multiple access schemes should consider individual UE's resource conditions and overall network performance. To the best of our knowledge, this important aspect has not been explored.
\par
Based on our literature review and analysis, the above works have two fundamental limitations: a). Most existing studies focus on simply increasing the overall spectral/energy efficiency to a higher level without considering UE-specific resource conditions, hardware constraints, and QoS demands. b). These works only utilize one or two radio resource domains without fully exploring all available multi-dimensional resources in improving the cost-effectiveness of multiple access.

\subsection{Contributions}
In achieving individualized QoS provisioning in 6G, we propose a flexible MDMA scheme in this paper, which can be viewed as a converged multiple access technique of OMA, power-domain NOMA, and spatial-domain NOMA. 
For a given UE, our proposed scheme can determine the most advantageous and suitable multiple access mode for it based on the evaluation of its distinctive radio resource utilization costs, conditions, and constraints in multiple domains. 
Specifically, the multi-dimensional resource utilization cost is defined to reflect the power consumption and complexity at UE side incurred by non-orthogonal interference cancellation of spatial- and power-domain NOMA. The main technical contributions of this paper can be summarized as follows:
\begin{itemize}
  \item Proposal of MDMA for individualized service provisioning while considering resource utilization cost. The newly developed MDMA scheme can be used to choose a multiple access mode (i.e., a cost-effective way to explore additional degrees of freedom in the multiple resource domains) for each UE that can strike a balance between its resource utilization cost and the performance gain of fulfilling its specific QoS demands. In detail, it consists of two stages: i.e., cost-aware selection of multiple access mode and situation-aware multi-dimensional radio resource allocation of users. Firstly, the base station (BS) adaptively organizes UEs with disparate multi-domain resource constraints as a coalition to reduce potential utilization costs. Then, multi-dimensional resource allocation is achieved by maximizing the sum of UEs' utility functions under individual UE's resource constraints and QoS requirements.
  \item Formation of UE coalitions to reduce UE-specific radio resource utilization cost while fully utilizing the available multi-dimensional radio resources. According to UE’s individual preference of resource and associated utilization cost, BS organizes mutually beneficial UEs into cooperative coalitions, that is, coexisting UEs can be multiplexed in any combinations of multi-dimensional resources with suitable multiple access mode to achieve a low resource utilization cost. The UE coalition formation algorithm is designed based on the two-sided many-to-one matching theory. The proposed algorithm can guarantee the exchange stability and only operates at a coarse time granularity, keeping UE coalitions unchanged for a relatively long period and thus reducing complexity.
  \item Management of multi-dimensional resource to meet UE-specific QoS requirements. Solving the resource allocation problem for the proposed MDMA scheme is non-convex and NP-hard, which is computationally intractable. This paper applies the successive convex approximation method to transform the original resource allocation problem into a concave optimization problem, for which existing convex optimization-based approaches can solve. Then, using the Lagrange dual decomposition method, this problem can be decomposed into independent sub-problems with closed-form solutions. Moreover, an iterative algorithm is proposed to further reduce the computational complexity.
\end{itemize} 
\par
The rest of the paper is organized as follows: Section~\ref{System Model} describes the system model employing MDMA schemes, then introduces the corresponding problem formulation. Section~\ref{Proposed Solutions} presents several adopted strategies to solve the non-convex utility sum maximization optimization problem. Finally, Section~\ref{Simulation} presents the simulation results and Section~\ref{conclusion} concludes this paper.

\section{System Model and Problem Formulation}\label{System Model}
The purpose of this paper is to design a MDMA scheme that can flexibly and opportunistically orchestrate multi-dimensional radio resources to meet each UE's specific service demands cost-effectively. In this section, to capture the influence of UE's specific resource situations on the MDMA, we will incorporate three essential factors into the system model, that is, the disparate resource conditions, heterogeneous resource utilization cost, and individualized resource constraints of each UE. In detail, Section \ref{CH_MODEL} (channel model) is set up to reflect the resource availability and constraint in the multi-dimensional radio resource domains. Then, Section \ref{MDMA_protocol} (cost-aware MDMA for individualized service provisioning) captures the impact of UE devices' radio resource utilization cost and hardware constraints on selecting the multi-dimensional multiple access modes. Based on the established system model, we introduce the proposed MDMA scheme and corresponding problem formulation.
\par
\begin{figure}[h]
    \vspace{-0.5cm}
    \centering
    \includegraphics[scale = 0.75]{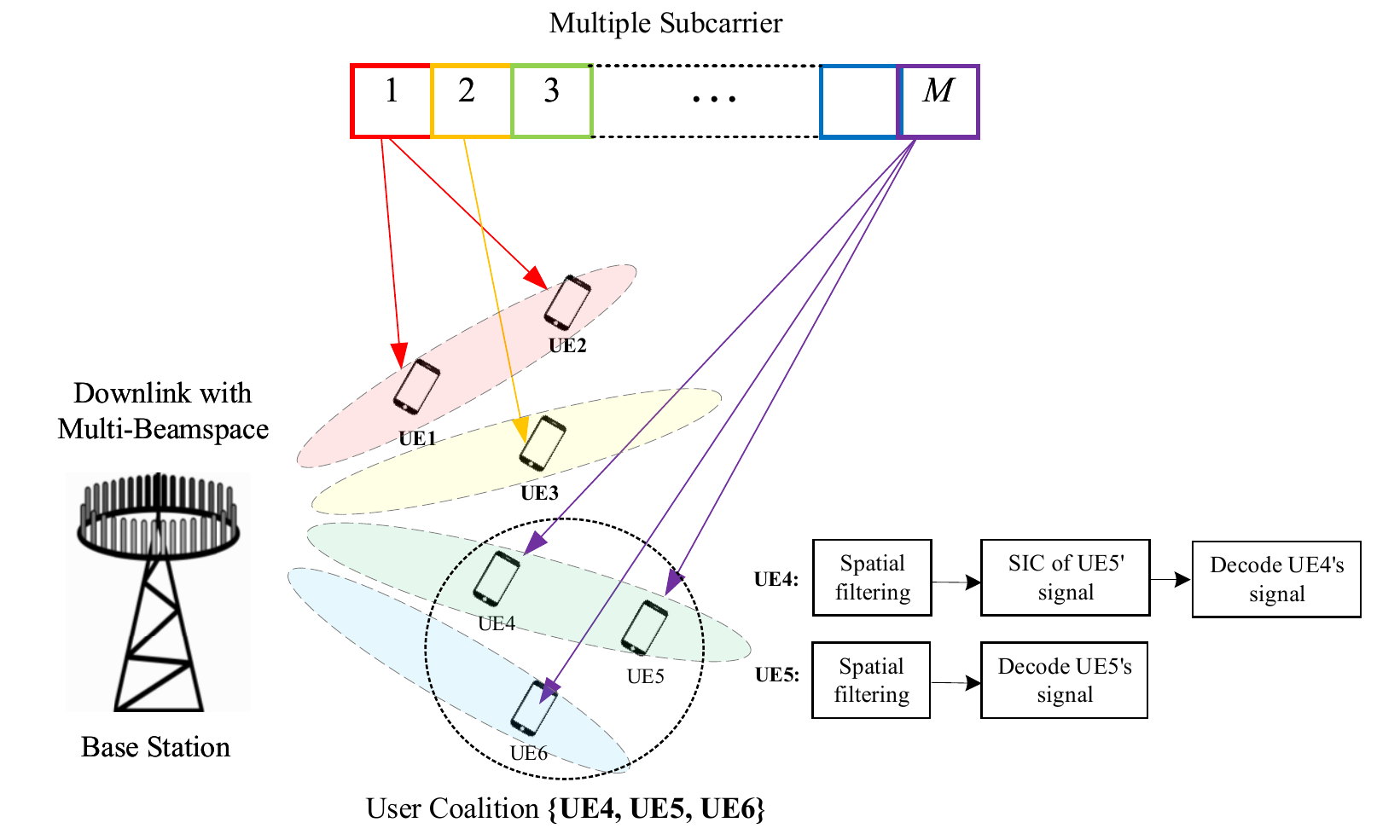}
    \caption{Illustration of proposed multi-dimensional multiple access (MDMA) scheme, which can flexibly utilize multi-dimensional radio resources by forming UE coalitions. For instance, ${\rm UE}4,\;{\rm UE}5$ and ${\rm UE}6$ have disparate multi-domain resource constraints: i) ${\rm UE}4$ and ${\rm UE}5$ have high channel gain difference in power domain but high spatial correlation; ii) ${\rm UE}6$ has good orthogonality with ${\rm UE}4$ and ${\rm UE}5$ in spatial domain. Thus, BS organizes these three UEs as a user coalition to share multi-dimensional resources cost-effectively, in which ${\rm UE}4$ and ${\rm UE}5$ are served by power-domain NOMA while ${\rm UE}6$ is served by beamforming.
    }\centering
    \vspace{-0.3cm}
    \label{1}
\end{figure}

Consider the downlink scenario in a cellular network shown in Figure~\ref{1}, where a single BS is deployed. Specially, the BS is located at the origin of a disk with radius $R$ and the BS is equipped with a uniform linear array, which has $N_{t}$ antennas. 
The total available bandwidth $B$ is divided into $M$ orthogonal subchannels (SCs). Meanwhile, the BS serves a set $\mathcal{K}$ of $K$ single-antenna UEs ($K>M$). For simplification, each UE only needs one SC to transmit. 

\subsection{Channel Model}\label{CH_MODEL}
As shown in Figure~1, the location of UE $k$ is characterized by $(d_{k}, \theta_{k})$, where $d_{k}$ is the distance between $u_{k}$ and the BS, and $\theta_k\in\left(-\pi,\pi\right)$ is the angles of departure (AOD) of UE $k$, seen from the broadside direction of the transmit antenna array, i.e., the physical direction of the line-of-sight (LOS) path. The channel vector with complex coefficients between the BS and UE $k$ in the $c$-th beamspace on $m$-th SC is defined as
\begin{small}
\begin{equation}\label{CH}
    \boldsymbol{h}_{k,m}=\sqrt{{\rm{PL}}(d_{k})}\cdot\boldsymbol{g}_{k,m}\in\mathbb{C}^{N_t\times 1,
    k \in {\mathcal{K}}},
\end{equation}
\end{small}where \begin{footnotesize}${\rm{PL}}(d_{k})$\end{footnotesize} denotes the large-scale fading from BS to the UE $k$. Furthermore, this paper assumes that the LOS path exists in the intra-cell communication links, then vector \begin{small}$\boldsymbol{g}_{k,m}$\end{small} follows uncorrelated Rician fading~\cite{8552437}, that is, 
\begin{small}
\begin{equation}
    \boldsymbol{g}_{k,m}=\sqrt {\frac{\kappa }{{\kappa  + 1}}}\cdot\boldsymbol{a}\left(\theta_k\right)+\sqrt {\frac{1 }{{\kappa  + 1}}}\cdot\boldsymbol{z}_{k,m}, k \in {\mathcal{K}}, \tag{\ref{CH}a}
\end{equation}
\end{small}where vector \begin{small}$\boldsymbol{a}\left(\theta_k\right)=\left[1,e^{{-j2{\pi}\Delta\sin(\theta_k)} },\cdots,e^{{-j2{\pi}\left(N_t-1\right)\Delta\sin(\theta_k)} }\right]$
\end{small} is accounting for the LOS component; $\Delta$ is the inter-antenna spacing in the unit of carrier wavelength, and vector \begin{small}$\boldsymbol{z}_{k,m}\sim\mathcal{CN}\left(\boldsymbol{0}_{N_t},\boldsymbol{I}_{N_t}\right)$\end{small} follows i.i.d. complex Gaussian distribution. 
\par
As a starting point, the spatial domain is coarsely divided into $B$ beamspaces according to the AoD of UEs (i.e., $\theta_{k}$) \cite{Wudan}. The set of UEs associated with the $b$-th beamspace is denoted as \begin{small}$\mathcal{B}_{b}$\end{small}, where \begin{small}$\bigcup_{b=1}^{B}{\mathcal{B}_{b}}=\mathcal{K}$\end{small} and \begin{small}$\mathcal{B}_{b}\cap\mathcal{B}_{b'}={\varnothing},\;\forall b,b'\in\mathcal{B}$\end{small}. The UEs in different beamspaces have enough spatial orthogonality. In contrast, the UEs within one beamspace cannot sufficiently guarantee spatial orthogonality. In this paper, \textit{multiple UEs in different beamspaces can share the same SC by spatial-domain NOMA (beamforming), while two UEs in the same beamspace can only share the same SC by power-domain NOMA}.
\subsection{Cost-Aware MDMA for Individualized Service Provisioning}\label{MDMA_protocol}
\subsubsection{Proposed MDMA scheme}

In this part, we design a MDMA scheme,in which BS adaptively choose a suitable multiple access mode for each UE based on their resource conditions, constraints, QoS demands, as well as utilization costs.
\par
\begin{figure}[!h]
    \vspace{-0.8cm}
    \centering
    \includegraphics[scale=0.4]{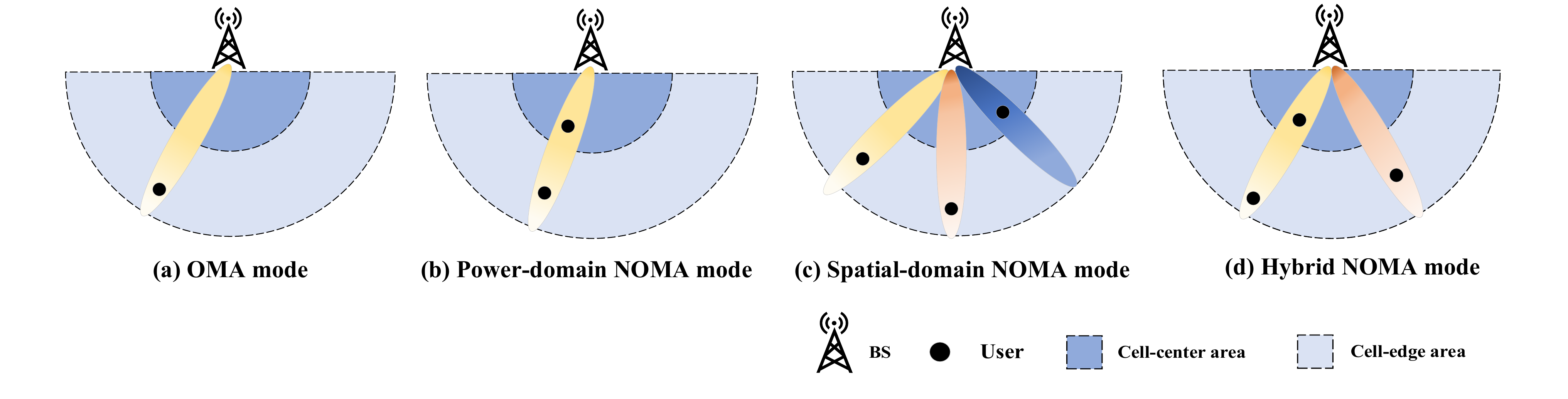}
    \caption{Candidate multiple access modes for UE coalition: Each UE can utilize the radio resource in OMA mode with very low utilization cost and high-quality links, when the network traffic is low. With the increasing of network traffic, orthogonal radio resources are insufficient, UEs with less resource sharing conflicts can be multiplexed by spatial and power domain NOMA.}\centering
    \label{Fig2}
    \vspace{-0.2cm}
\end{figure}

Firstly, this paper introduces the concept of UE coalition, that is, several UEs with disparate multi-dimensional resource constraints can coordinate as resource-sharing coalitions to utilize the same SC in a cost-effective and less conflicting way. As shown in Fig.~\ref{Fig2}, there are four multiple access modes for each UE, i.e., \textit{OMA mode}, \textit{power-domain NOMA mode}, \textit{spatial-domain NOMA mode}, and \textit{hybrid multiple access mode}. For instance, spatial-domain NOMA mode would be applied in one SC if UEs within the coalition have large separated AOD. Meanwhile, power-domain NOMA mode could be used to serve the near-UE and the far-UE with large channel gain difference. Furthermore, to maintain low SIC complexity, the number of users in each power-domain NOMA pair is limited to 2, which is a common assumption in the related works~\cite{ding2020unveiling}.
\begin{definition}[\textbf{UE Coalition}]\label{D1}
UEs with low intensity of radio resource sharing conflicts can form a coalition, which will share and utilize the same SC by preferred multiple access mode. All UEs are partitioned into a set \begin{small}$\mathcal{N}$\end{small} of $M$ coalitions according to the UE's resource utilization costs, hardware constraints, and perceived value of radio resources. Specifically, the $n$-th UE coalition is defined as a UE set \begin{small}${{\cal U}_n}$\end{small},
where \begin{small}$\cup_{n \in \mathcal{N}}{\mathcal{U}_{n}}=\mathcal{K}$\end{small} and \begin{small}$\cap_{n \in \mathcal{N}}{\mathcal{U}_{n}}={\varnothing}$\end{small}. In particular, to describe the employed multiple access mode of UEs in $n$-th coalition, a binary variable $\alpha_n$ to denote if spatial-domain NOMA is defined as,\begin{small}
\begin{numcases}
{{\alpha _n} = }
{1,\;{\rm{if}}\;\sum\nolimits_{b=1}^{B} {\boldsymbol{1}{\left\{ {{\left| {{{\mathcal{B}}_b} \cap {{\mathcal{U}}_n}} \right|}  \ge 1} \right\}} > 1,} \notag} \\
{0,\;{\rm{otherwise.}}\notag}
\end{numcases}\end{small}where \begin{small}$\boldsymbol{1}\left\{\cdot\right\}$\end{small} is the indicator function. Meanwhile, let $\beta_n=1$ denote if power-domain NOMA mode is used by UEs belonging to the same beamspace; otherwise, $\beta_n=0$.
\begin{small}
\begin{numcases}
{{\beta _n} = }
{1,\;{\rm{if}}\;\exists \;b{\rm{: }\;} {{\left| {{{\mathcal{B}}_b} \cap {{\mathcal{U}}_n}} \right|}}  = 2, 1 \le b \le B,\notag} \\
{0,\;{\rm{otherwise.}}\notag}
\end{numcases}
\end{small}Hence, there are four candidate multiple access modes for $n$-th UE coalition to multiplex the same SC,\begin{small}
\begin{numcases}
{\left( {{\alpha _n},{\beta _n}} \right)= } 
\left(0,0\right),\;{\rm{
if\;OMA\;is\;set}} \notag\\
{\left(1,0\right),\;{\rm{if\;spatial\;domain\;NOMA\;is\;set}}\notag} \\
{\left(0,1\right),\;{\rm{if\;power\;domain\;NOMA\;is\;set}}\notag} \\
{\left(1,1\right),\;{\rm{if\;hybrid\;NOMA\;mode\;is\;set}}\notag}
\end{numcases}
\end{small}
\end{definition}
\subsubsection{Signal Transmission Model}
Let $s_{n,m}$ denote the SC allocation indicator, where $s_{n,m}=1$, when $m$-th SC is utilized by UE coalition \begin{small}$n \in \mathcal{N}$\end{small}; otherwise,  $s_{n,m}=0$. Furthermore, each UE is assigned with a beamforming vector to exploit the gain in the spatial domain. Let $\boldsymbol{v}_{k}\in\mathbb{C}^{N_t\times1}$ denotes the beamforming vector from BS to UE $k$. Herein, zero-forcing beamforming is used to generate the beamforming vector and concentrate the signal power to every UE's direction. Then, the signal-to-interference-plus-noise rate (SINR) of UE $k \in {\mathcal{U}_n}$ on $m$-th SC is expressed as:
\begin{small}
\begin{align}\label{EQ_3}
&\gamma _{k,m}=\frac{p_{k}\left | \boldsymbol{h}^{\rm H}_{k,m}\boldsymbol{v}_{k}\right |^{2}}{ I_{k,m}+N_0},\;{\rm{if}}\;s_{n,m}=1,
\end{align}
\end{small}where $p_{k}$ denotes the downlink transmission power of $k$-th UE, term $N_0$ is the additive Gaussian noise on each SC, and term $I_{k,m}$ in the denominator of the SINR represents the non-orthogonal interference. As illustrated in Figure~\ref{Fig2}, there are four kinds of interference situations, i.e.,
\begin{small}
\begin{numcases}
{I_{k,m}=} 
0,\;{\rm{
if\;OMA\;mode,}} \notag\\
{I_{k,m}^{\rm{PD}},\;{\rm{if\;power\;domain\;NOMA\;mode,}}\notag} \\
{I_{k,m}^{\rm{SD}},\;{\rm{if\;spatial\;domain\;NOMA\;mode,}}\notag} \\
{I_{k,m}^{\rm{PD}} + I_{k,m}^{\rm{SD}},\;{\rm{if\;hybrid\;NOMA\;mode,}}\notag}
\end{numcases}
\end{small}In the case of the OMA mode, there is no additional interferences, i.e., $I_{k,m}=0$. In the case of power-domain NOMA mode, the SIC receiver of ``near-UE" $k$ can cancel the interference from ``far-UE" with lower channel gain. Then, the interference in power-domain (PD), \begin{small}$I_{k,m}^{\rm{PD}}$\end{small}, is
\begin{small}
\begin{align} 
I_{k,m}^{\rm{PD}}= 
\sum\nolimits_{b=1}^{B} \boldsymbol{1}{\left\{k\in\mathcal{B}_b\right\}} \cdot  \sum\nolimits_{i \in \mathcal{S}_{k}} p_i\left |\boldsymbol{h}^{\rm{H}}_{k,m}\boldsymbol{v}_{i}\right |^{2}, \tag{\ref{EQ_3}a}
\end{align}\end{small}where \begin{small}$\boldsymbol{1}\left\{\cdot\right\}$\end{small} is the indicator function and \begin{small}$\mathcal{S}_{k}=\left \{i|i\in {{\mathcal{U}_n}{\cap}{\mathcal{B}_b}}, \|\boldsymbol{h}_{i,m}\|_{2}>\|\boldsymbol{h}_{k,m}\|_{2} \right \}$\end{small} is the UE who has better channel gain than UE $k$. In the case of spatial-domain NOMA mode, \begin{small}$I_{k,m}^{\rm{SD}}$\end{small} is the non-orthogonal interference in the spatial-domain (SD), caused by the UEs in other beamspaces,\begin{small}
\begin{align} 
I_{k,m}^{\rm{SD}} = \sum\nolimits_{b=1}^{B} \boldsymbol{1}{\left\{k\notin\mathcal{B}_b\right\}} \cdot \sum\nolimits_{i \in {{\mathcal{U}_n}{\cap}{\mathcal{B}_b}}} p_i \left |\boldsymbol{h}^{\rm{H}}_{k,m}\boldsymbol{v}_{i}\right |^{2}, \tag{\ref{EQ_3}b}
\end{align}
\end{small}Moreover, in the case of the hybrid NOMA mode, it will cause the non-orthogonal interferences in both spatial and power domains.
\par
Therefore, the data rate of UE $k$ in $n$-th UE coalition can be expressed as,
\begin{small}
\begin{align}\label{data_rate}
r_{k}= \sum\nolimits_{m = 1}^{M} {s_{n,m}\cdot\frac{B}{M}\log_{2}\left ( 1+\gamma _{k,m}\right )},\;\;k\in\mathcal{U}_n.
\end{align}
\end{small}\subsubsection{Divergent Recourse Constraints Experienced by each UE}\label{Formulate_Problem}
The proposed MDMA scheme aims to encourage UEs to explore multi-dimensional radio resources through the selection of multiple access modes. However, the utilization of radio resources in different dimensions comes at different costs of computational complexity, subject to heterogeneous hardware constraints. Specifically, in NOMA modes, UE will suffer from the non-orthogonal interference caused by the partially overlapped signal in the spatial-domain, the power-domain, or both, which inevitably induces additional power consumption of UEs for interference mitigation. Therefore, we have the following definitions of the utilization costs and resource constraints at the UE side.
\par
\textit{i). Radio resource utilization cost of UE}. Firstly, the power-domain NOMA requires the SIC processing at the ``near-UE'' (with strong channel gain), which leads to extra complexity and power consumption at the receiver \cite{Liu_TWC}. The \textit{utilization cost to the power-domain NOMA} for the ``near-UE" is inversely proportional to the experienced SINR in the SIC prodecure \cite{PD_NOMA_cost}. For UE $k \in \mathcal{U}_n$ on $m$-th SC, if UE $k$ has strong channel gain in the power-domain NOMA pair, its utilization cost for SIC is defined as\begin{small}
\begin{align} 
    &\psi _{k,m}^{{\rm{PD}}} =   
    \sum\nolimits_{b = 1}^{B} \boldsymbol{1}{\left\{k \in \mathcal{B}_b\right\}} \cdot \sum\nolimits_{i \in {\mathcal{X}_k}} {\left[{\rho_{0}}-{\rho_{1}\lg\left(\gamma_{k,m}^{\rm sic}\right)}\right]}
    ,\;{\rm{if}}\;s_{n,m} = 1\;{\rm and}\; \beta_n=1.
\end{align}
\end{small}where term $\rho_{0}$ denotes the constant cost of SIC processing, \begin{small}$\mathcal{X}_{k}=\left \{i|i\in {{\mathcal{U}_n}{\cap}{\mathcal{B}_b}}, \|\boldsymbol{h}_{k,m}\|_{2}>\|\boldsymbol{h}_{i,m}\|_{2} \right \}$\end{small} is the UE who has worse channel gain than UE $k$, ${\rho}_{1}$ is the positive scalar, and $\gamma_{k,m}^{\rm sic}$ denotes the SINR experienced by UE $k$, when UE $k$ detects the signal of UE $i$, i.e., ``far-UE", in presence of interference from its desired signal, that is,\begin{small}
\begin{equation}
    \gamma_{k,m}^{\rm sic} =\frac{p_{i} | \boldsymbol{h}^{\rm H}_{k,m}\boldsymbol{v}_{i} |^{2}}{ p_{k} | \boldsymbol{h}^{\rm H}_{k,m}\boldsymbol{v}_{k}|^{2} + I_{k,m}^{\rm SD}+N_0}.\notag
\end{equation}
\end{small}As we can see, $\psi _{k,m}^{{\rm{PD}}}$ is an increasing function of the inverse SINR, i.e., $(\gamma_{k,m}^{\rm sic})^{-1}$. Besides, only the UE with strong channel gain has the additional cost by the SIC processing. Thus, the far-UE, which does not perform SIC, has no additional utilization cost in the power domain.
\par
Secondly, for the case of spatial-domain NOMA, the high spatial correlation between desired signal and interference signal makes it costly for the UE side to distinguish the overlapped signals in the spatial domain~\cite{JSAC_Yanan}. The \textit{utilization cost of the spatial domain} is determined by the spatial correlation among UEs in different beamspaces sharing the same SC. For UE $k \in {\mathcal{U}_n}$ on $m$-th SC, the spatial domain non-orthogonality of this UE is defined as\begin{small}
\begin{equation}
    \psi _{k,m}^{{\rm{SD}}} = \sum\nolimits_{b = 1}^{B} \boldsymbol{1}{\left\{k\notin\mathcal{B}_b\right\}} \cdot {\sum\nolimits_{i \in {{\mathcal{U}_n} {\cap} \mathcal{B}_b}} { \rho_{2} \cdot  {\frac{{\left| {{{\boldsymbol h}_{k,m}^{\rm H}} {\boldsymbol h}_{i,m}} \right|}}{{\| {{{\boldsymbol h}_{k,m}}} \|_2 \cdot \| {{{\boldsymbol h}_{i,m}}} \|_{2}}}} }},\;{\rm{if}}\;s_{n,m} = 1\;{\rm and}\;\alpha_n =1
\end{equation}
\end{small}where term ${\rho}_{2}$ is a positive scalar related to the additional costs of using spatial domain NOMA at the receiver. Then, we quantify the utilization costs of radio resource with respect to the non-orthogonality in both power and spatial domain,
\begin{definition}[\textbf{Multi-dimensional Radio Resource Utilization Cost at UE side}]\label{D4}
In UE coalition \begin{small}$n \in \mathcal{N}$\end{small}, the corresponding utilization cost of UE \begin{small}$k \in {\mathcal{U}_n}$\end{small} is defined as
\begin{small}
\begin{align} \label{cost_formula}
    g_{k} &= \sum\nolimits_{m = 1}^{M} s_{n,m} \cdot g_{k,m} = \sum\nolimits_{m = 1}^{M} s_{n,m} \cdot \left({\alpha_m \cdot \psi _{k,m}^{{\rm{PD}}}} + { \beta_m \cdot {\psi _{k,m}^{{\rm{SD}}}}}\right),
\end{align}
\end{small}If coalition $n$ utilizes $m$-th SC by OMA mode, the utilization cost of UE \begin{small}$k \in {\mathcal{U}_n}$\end{small} is zero.
\end{definition}
\par
\textit{ii). Hardware Constraints of UE.} Ideally, all UEs are expected to be equipped with compatible processing capabilities and functionalities for different multiple access modes. However, in practice, different UE has diversified processing capabilities and limitations, which may restrict UE's selection on specific multiple access modes. Particularly, power-domain NOMA requires ``near-UE" (i.e. UE with strong channel gain) to perform SIC. However, the SIC capability is not universally exist considering the heterogeneity of device type.
\par
 In this paper, assume that the UEs with SIC capability as set $\mathcal{K}_{\rm sic}$ and the UEs without SIC capability as $\mathcal{K}_{\rm no-sic}$, where ${\mathcal{K}_{\rm sic}} \cup {\mathcal{K}_{\rm no-sic}} = \mathcal{K},{\mathcal{K}_{\rm sic}} \cap {\mathcal{K}_{\rm no-sic}} = {\varnothing}$. If UE $k$ in $n$-th UE coalition does not possess SIC ability, it cannot be selected as the ``near-UE" in the power-domain NOMA pair. This hardware constraint can be mathematically expressed as
\begin{small}
\begin{equation}\label{C1}
    \sum\nolimits_{k \in {{\mathcal{U}_n}}} \sum\nolimits_{b = 1}^{B} \sum\nolimits_{i \in {{\mathcal{U}_n}{\cap}{\mathcal{B}_b}}}  \boldsymbol{1}{\left\{k\in{\mathcal{K}_{\rm no-sic}}\right\}} \cdot \boldsymbol{1}{\left\{{\|{\boldsymbol{h}}_{k,m}\|}_2 > {\|{\boldsymbol{h}}_{i,m}\|}_2\right\}} = 0,\;\forall n \in  \mathcal{N}. \tag{C1}
\end{equation}
\end{small}\subsection{Problem Formulation}
\subsubsection{UE's perceived value of radio resource} In principle, we aim at jointly balancing two conflicting metrics for each UE, that is, a) QoS performance and b) utilization costs of multi-dimensional radio resources. For UE $k \in \mathcal{U}_{n}$, its cost-aware utility function is defined as\begin{small}
\begin{equation}\label{U_f}
    {u_{k}} = r_{k}/R_{\max} - w_k \cdot {g_{k}},\;k \in \mathcal{U}_{n},
\end{equation}
\end{small}where $R_{\max}$ is the ideal data rate of UE $k$ and ${w}_{k}$ is the weighting factor to normalize the units of two metrics. For instance, if one UE is energy-sensitive (e.g. limited battery), its $w_{k}$ will be set to a high value to restrict the energy consumption at the receiver side, and vice versa.\\
\textbf{\textit{Remark}}: It should be noted that the same SC might worth different utility values, i.e., formula (\ref{U_f}), to different UE of different coalitions. By exploiting this feature, we can utilize multi-dimensional radio resources more efficiently and opportunistically for individualized QoS provisioning. 
\subsubsection{Target Problem}
From the service-provisioning perspective, the proposed MDMA scheme should satisfy the distinctive recourse constraints and QoS requirements from each UE. In this paper, the overall objective is to maximize the total sum of UE's utility function subject to the hardware and resource constraints of UE. Thus, it can be formulated as a maximization optimization problem,\begin{small}
\begin{align}
&\mathcal{P}:\;\;\mathop{\max}\limits_{\Omega}  \left\{ { \sum\nolimits_{n \in \mathcal{N}} \sum\nolimits_{k \in \mathcal{U}_{n}} {{u_k}} } \right\} \notag\\
&\text{s.t.:}\;\;{\rm{C1}}: \sum\nolimits_{k \in {{\mathcal{U}_n}}} \sum\nolimits_{b = 1}^{B} \sum\nolimits_{i \in {{\mathcal{U}_n}{\cap}{\mathcal{B}_b}}}  \boldsymbol{1}{\left\{k\in{\mathcal{K}_{\rm no-sic}}\right\}} \cdot \boldsymbol{1}{\left\{{\|{\boldsymbol{h}}_{k,m}\|}_2 > {\|{\boldsymbol{h}}_{i,m}\|}_2\right\}} = 0,\forall n \in \mathcal{N},\nonumber\\ 
&\;\;\;\;\;\;\;{\rm{C2}}: 1 \le \left|\mathcal{U}_{n}\right| \leq L_{\max},\cup_{n=1}^{M}{\mathcal{U}_{n}}=\mathcal{K}, \cap_{n=1}^{M}{\mathcal{U}_{n}}={\varnothing},  \nonumber\\
&\;\;\;\;\;\;\;{\rm{C3}}: \sum\nolimits_{m=1}^{M} s_{n,m}=1, s_{n,m}=\left \{ 0,1\right \}, \forall n\in\mathcal{N},\nonumber\\
&\;\;\;\;\;\;\;{\rm{C4}}: R_{\rm{th}} \le {r_{k}} \le R_{\max }, \forall k \in \mathcal{K}, \nonumber\\
&\;\;\;\;\;\;\;{\rm{C5}}: \sum\nolimits_{ k \in \mathcal{K}}p_{k} \leq P_{\max} ,\nonumber
\end{align}\end{small}where \begin{small}
$\Omega=\left\{(\mathcal{U}_n, \alpha_n, \beta_n), s_{n,m}, p_{k}\right\}$\end{small} is \textit{UE coalition formation, multiple access mode selection, subchannel, and power allocation}. Constraint ${\rm C}1$ is the hardware constraints of UE. Constraint ${\rm C}2$ restricts the UE coalition, that every coalition can have at lest one UE and at most $L_{\max}$ UEs. Constraint ${\rm C}3$ allows every UE coalition is allocated to one SC. Constraint ${\rm C}4$ ensures that data rate of UE $k$ is between the minimum data rate for requirement satisfaction (i.e., $R_{\rm{th}}$) and the the ideal data rate (i.e. $R_{\max}$). ${\rm C}5$ ensures the total transmit power will not exceed maximum power of BS.  
\section{Proposed Solution for Target Problem}\label{Proposed Solutions}

The target problem $\mathcal{P}$ is NP-hard due to its mixed-integer objective function and complicated constraints. To deal with its NP-hardness, problem $\mathcal{P}$ is divided into two sub-problems, which are solved separately in two stages. Firstly, based on the aforementioned MDMA scheme, heterogeneous UEs with diverse QoS requirements and inherent hardware constraints are coordinated to share and utilize the same SC in a flexible way, which can greatly improve overall resource utilization efficiency. However, UEs who share the same SC  suffer from a certain amount of non-orthogonal interference due to the existence of resource sharing conflicts. Therefore, we initiate problem-solving strategy by creating UE coalitions with consideration of UEs' location information, which aims to group multiple UEs with acceptable intensity of radio resource sharing conflicts as a coalition. Secondly, in stage II, according to real-time CSI, we jointly allocate SC and power to each UE coalition to maximize the total sum of UE's utility function.
\subsection{Stage I--UE Coalition Formation}\label{coalition}
 {By building a mutually reciprocal paradigm, UEs in the same coalition will not only meet individual QoS requirements, but more importantly, yield acceptable radio resource utilization costs with suitable multiple access modes. Thus, the UE coalition formation problem is formulated as}, \begin{small}
\begin{align}
&\mathcal{P}_1:\;\;\mathop{\min}\limits_{\left\{\mathcal{U}_n, \alpha_n, \beta_n\right\}}  \left\{ { \sum\nolimits_{n \in \mathcal{N}} \sum\nolimits_{k \in \mathcal{U}_{n}} {w_k \cdot {{\tilde g}}_k^n} } \right\} \notag \\
&\text{s.t.:}\;\;{\rm{{\tilde C1}}}: \sum\nolimits_{k \in {{\mathcal{U}_n}}} \sum\nolimits_{b = 1}^{B} \sum\nolimits_{i \in {{\mathcal{U}_n}{\cap}{\mathcal{B}_b}}}  \boldsymbol{1}{\left\{k\in{\mathcal{K}_{\rm no-sic}}\right\}} \cdot \boldsymbol{1}{\left\{{\rm PL}(d_k) > {\rm PL}(d_i)\right\}} = 0,\forall n \in \mathcal{N},{\rm{C2}}, \notag 
\end{align}\end{small}where \begin{small}${{{\tilde g}}_k^n}$\end{small} represents the intensity of radio resource sharing conflict experienced by UE $k$ when it joins coalition \begin{small}$n\in\mathcal{N}$\end{small}. Specifically, the intensity of radio resource sharing conflict of UE \begin{small}$k \in \mathcal{U}_n$\end{small} is defined as an approximation of radio resource utilization cost (\ref{cost_formula}) in \textit{\textbf{Definition \ref{D4}}}, that is, \begin{small}
\begin{align}
    {{{\tilde g}}_k^n} &= \rho_{0} - {{\rho_{1}} \cdot {\sum\nolimits_{b = 1}^{B} \boldsymbol{1}{\left\{k\in\mathcal{B}_b\right\}} \cdot \sum\nolimits_{i \in {{\mathcal{U}_n}{\cap}{\mathcal{B}_b}}}  \boldsymbol{1}{\left\{{\rm PL}(d_k) > {\rm PL}(d_i)\right\}} \cdot 
    \lg\left[\frac{{\rm{PL}}(d_{k})}{{\rm{PL}}(d_{i})+N_0/{P_{\max}}}\right]}} \notag \\ 
    &\;\;\;\;\;\;\;\;\;\;\;\;\;\;\;\;\;\;\;\;\;\;\;\;\;\;\;\;\;\;\;\;\;\;\;\;\;\;\;\;\;\;\;\;\;\;\;\;\;\quad\quad\;\;
    + {{\rho_{2}} \cdot {\sum\nolimits_{b = 1}^{B} \boldsymbol{1}{\left\{k\notin\mathcal{B}_b\right\}} \cdot {\sum\nolimits_{i \in {{\mathcal{U}_n} {\cap} \mathcal{B}_b}} {\left| {{\boldsymbol{a}\left(\theta_k\right)^{\rm H}} \boldsymbol{a}\left(\theta_i\right)} \right|}}}},
\end{align}\label{matching preference}
\end{small}which is only based on location information of UE $\left(d_k,\theta_k\right)$. \\
\textbf{\textit{Remark}}: Since the location information of UE typically changes at a slow timescale (e.g., tens of milliseconds), the UE coalition formation process can be executed in a coarse time granularity.
\par
\subsubsection{Matching Problem Formulation for UE Coalitions}
In this part, a set of coalitions that may differ in terms of their capacity and multiple access mode will be formed by various UEs. Each UE can be allocated to only one coalition of which the resultant resource sharing conflict and performance gains should satisfy individual requirements. Hence,
The UE coalition formation process in problem $\mathcal{P}_1$ can be treated as a multi-dimensional stable roommate (MD-SR) problem, where individual \textit{agents} have to be allocated to dynamically sized \textit{rooms} and have diversity preferences over their potential roommates \cite{boehmer2020stable}. 
In this respect, matching theory demonstrates its effectiveness when coping with heterogeneous UEs, each of which has its own type, objective, and constraint.
To elaborate, the two disjoint sets of UEs (\begin{small}$\mathcal{K}$\end{small}) and coalitions (\begin{small}$\mathcal{N}$\end{small}) are considered as \textit{agents} and \textit{rooms}. Each UE (agent) chooses to join coalition (room) according to their diverse \textbf{tolerance} on the radio resource sharing conflict in each coalition. As formulated in problem $\mathcal{P}_1$, each UE tends to select and form the coalition which brings about the lowest radio resource sharing conflict. Before solving problem $\mathcal{P}_1$, we first present the definition of matching as below:
\begin{definition}[\textbf{Coalition Formation}]\label{D3}
 The coalition formation can be achieved and expressed by a two-sided many-to-one matching $\Pi$. Mathematically, matching $\Pi$ is a partition of \begin{small}$\mathcal{K}$\end{small} into $N$ disjoint un-ordered sets, i.e., the UE coalitions \begin{small}$\mathcal{N}=\{\mathcal{U}_1,\mathcal{U}_2,\dots,\mathcal{U}_n\}$\end{small}.  Let \begin{small}$\Pi(k)$\end{small} denote UE $k$'s matched coalition in $\Pi$, then we have \begin{small}$\Pi(k)\subseteq\mathcal{N}, \forall k\in\mathcal{K}$\end{small} and \begin{small}$\Pi(n)\subseteq\mathcal{K}, \forall n\in\mathcal{N}$\end{small}. The cardinality of coalition \begin{small}${{\cal U}_n}$\end{small}, \begin{small}$|\mathcal{U}_n|$\end{small}, is endowed with \begin{small}$l \in \mathbb{N}^{+}$\end{small}, satisfying \begin{small}$1\leq l\leq L_{\max}$\end{small}. 
\end{definition}
\par
During the coalition formation process, we use \textit{preference relation} to describe the \textit{interactions} between UEs and coalitions. As mentioned before, UEs in the same coalition will have radio resource sharing conflicts that induce \textit{radio resource utilization cost} in \textit{\textbf{Definition}} \textbf{\textit{\ref{D4}}}. Each UE \begin{small}$k\in \mathcal{K}$\end{small} uses \textit{preference relation} $\succ_{k}$ to rank the matched coalitions in strict order of resource-sharing conflicts. Therefore, UEs aim to choose certain peers that will introduce the least inherent resource sharing conflicts to form the coalition.
On the other hand, constraint ${\rm{{\tilde C1}}}$ and ${\rm{{C2}}}$ in $\mathcal{P}_1$ restrict the coalition formation process. If a group of UEs violate ${\rm{{\tilde C1}}}$ or ${\rm{{C2}}}$, they are not allowed to form a UE coalition and defined as \textit{forbidden coalition}. In detail, a binary variable $x_{n}$ is used to denote the feasibility and validity of coalition $n$, $x_{n}=1$ indicates $n$-th coalition violates constraint ${\rm{{\tilde C1}}}$ or ${\rm{{C2}}}$. 
Considering the cases of \textit{forbidden coalition}, UE $k$ prefers coalition $n$ to another coalition $n'$, if \begin{small}${{{\tilde g}}_k^n} < {{{\tilde g}}_k^{n'}}$\end{small} and $x_n=0$ are simultaneously satisfied, that is, \begin{small}\begin{equation}
    n\succ_{k}n' \Leftrightarrow  {{{\tilde g}}_k^n} < {{{\tilde g}}_k^{n'}}\;{\rm and}\;x_n=0,\; k \in \mathcal{K}. \notag
\end{equation}\end{small}The corresponding MD-SR problem $\mathcal{P}_1$ is proved to be NP-complete \cite{8675293}. Hence, a heuristic two-phase matching algorithm has been adopted to obtain a feasible coalition formation and achieve exchange stability.
\subsubsection{Matching Algorithm Design for Coalition Formation}\label{matching}
Due to the coexistence of diversity preference and forbidden coalitions, the objectives of matching algorithm design will not only focus on the establishment of a feasible solution, but also on the realization of matching stability and optimality. Hence, the first phase in the designed algorithm aims to obtain a feasible and valid coalition formation $\Pi$, i.e., there is no forbidden coalitions and all UEs/coalitions are matched.  
\par 
\textit{a) Phase I}: Firstly, according to the preference relation, UEs will compete to monopolize each coalition which always leads to OMA mode with zero resource sharing conflict. However, since the number of all UEs, $K$, is larger than the number of coalitions, $N$, this will make certain UEs remain unmatched and such coalition formation results cannot be validated. Denote \begin{small}$\mathcal{N}_{\mathrm{empty}}\subseteq\mathcal{N}$\end{small} as the set of empty coalitions that have not been populated by any UEs, and \begin{small}$\mathcal{N}_{\mathrm{C}}$\end{small} as the complementary set of \begin{small}$\mathcal{N}_{\mathrm{empty}}$\end{small}, where \begin{small}
$\mathcal{N}_{\mathrm{empty}}\cap{\mathcal{N}_{\mathrm{C}}}=\mathcal{N}$\end{small}.
Initially, UEs will be assigned to a random coalition available in \begin{small}$\mathcal{N}_{\mathrm{empty}}$\end{small} if \begin{small}$\mathcal{N}_{\mathrm{empty}}\neq\varnothing$\end{small}.
Otherwise, UEs will have to search for the coalitions that occupied by other UEs, and pick the best one that can bring lowest radio resource sharing conflicts with no \textit{forbidden coalition},\begin{small}
\begin{equation}\label{EQ-9}
    \mathcal{N}_{n^*} = \arg \min\nolimits_{\left\{n\in\mathcal{N}_{\mathrm{C}}\right\}} {\left\{w_k \cdot {{\tilde g}}_k^n\right\}},\; x_{n^*}=0.
\end{equation}
\end{small}This phase terminates when all UEs and coalitions are matched. Both the obtained coalition formation $\Pi$ and multiple access mode parameters \begin{small}$(\alpha_n,\beta_n)$\end{small} for each \begin{small}${{\cal U}_n}$\end{small} will be returned to the next phase.  
\par 
\textit{b) Phase II}: Such obtained matching result is unstable as some UEs may still prefer other coalitions to their current \textbf{matched} peers. We then utilize the inter-dependency of each UE's selection over individual coalitions to further optimize and stabilize the coalition formation result. Specifically, a \textit{rotation sequence} is adopted here with two major purposes. \\
i) Select a certain number of UEs and exchange their coalitions in a \textit{cyclic-shift pattern}. Namely, we \textit{rotate} some UEs' matching relations with their current coalitions and generate new coalition:
\begin{definition}[\textbf{Rotation Sequence}]\label{D6}
Rotation sequence $\xi_s$ is defined by a subset of UEs \begin{small}$\mathcal{K}_r\subseteq \mathcal{K}$\end{small} and current coalition formation $\Pi$ ,\begin{small}\begin{equation}
    \xi_{s}=\left\{\left[{\mathcal{K}_{r,1},\Pi\left(\mathcal{K}_{r,s+1}\right)}\right], \left[{\mathcal{K}_{r,2},\Pi\left(\mathcal{K}_{r,s+2}\right)}\right],\cdots,\left[\mathcal{K}_{r,S},\Pi\left(\mathcal{K}_{r,s}\right)\right]\right\},\;1\leq s \leq S-1, \notag
\end{equation}\end{small}where \begin{small}$\mathcal{K}_r=\{\mathcal{K}_{r,1},\mathcal{K}_{r,2},\dots,\mathcal{K}_{r,S}\}$\end{small}, \begin{small}$\left|\mathcal{K}_r\right|=S \geq 2$\end{small} indicating that there are a total of \begin{small}$S$\end{small} selected UEs in the subset, and $s$ is the index in rotation sequence $\xi_s$ used to define the re-matching relations. For instance, $\xi_s$ means that UE of \begin{small}$\mathcal{K}_{r,1}$\end{small} is rotationally shifted to its $s$-th contiguous neighbor's coalition \begin{small}$\Pi\left(\mathcal{K}_{r,s+1}\right)$\end{small}.
\end{definition}
ii) Create traversal for the rotated \textbf{coalition formation results} as an endeavor to reduce the total weighted resource sharing conflicts among all UEs, i.e., \begin{small}${ \sum\nolimits_{n \in \mathcal{N}} \sum\nolimits_{k \in \mathcal{U}_{n}} {w_k \cdot {{\tilde g}}_k^n} }$\end{small}. In detail, given a particular UE subset $\mathcal{K}_r$ with length of \begin{small}$S$\end{small}, there will be at most \begin{small}$S-1$\end{small} possible valid \textbf{coalition formations} in contrast to the original one $\Pi$. Denote the rotated new coalition result as \begin{small}$\Pi_{\mathcal{K}_r,\xi_s}$\end{small} and the case of \begin{small}$s=S$\end{small} will not be considered since \begin{small}$\Pi_{\mathcal{K}_r,\xi_s}$\end{small} will become the same as the original $\Pi$. Hence, we can take advantage of the rotation sequence to reach optimality in an iterative way: for each selected $\mathcal{K}_r$ and $\xi_s$, we compare the  weighted resource sharing conflict sum of each valid coalition formation and keep record of the best performance, which is given by \begin{small}\begin{equation}
    \Pi_{\mathcal{K}^{*}_r,\xi_{s^*}}=\arg \min\nolimits_{\left\{1\leq s \leq S-1\right\}} {\sum\nolimits_{n \in \mathcal{N}}\sum\nolimits_{k \in \mathcal{U}_{n'}}\left\{w_k \cdot {{\tilde g}}_k^n\right\}}. \notag
\end{equation}
\end{small}This process will continue until we have traversed all rotation sequences with no further improvement. The final coalition formation result in the rotation phase should converge to a local optimal point such that stability is achieved. This prodecure is presented in \textbf{Algorithm \ref{alg1}}.
\begin{proposition}
\textbf{Algorithm \ref{alg1}} can guarantee the exchange stability \cite{2016transfers} and converge to the local minimum point of NP-hard problem $\mathcal{P}_1$ by the operations defined by rotation sequences.   
\end{proposition}
\begin{IEEEproof} The detailed proof is discussed in the \textbf{Appendix A}.
\end{IEEEproof}
\vspace{-0.3 cm}
\begin{algorithm}
    \setstretch{1} 
    \caption{Coalitional Matching Algorithm}
    \label{alg1}
    \begin{scriptsize}
    	\SetKwInOut{Input}{Input}\SetKwInOut{Output}{Output}
        \Input{The set of UEs $\mathcal{K}$, and set of coalitions $\mathcal{N}$.} 
        \Output{A stable matching $\Pi^*$ for UE coalitions, multiple access mode indicators $\{\alpha_n^*, \beta_n^*\}$ for each $\mathcal{U}_n$, $n \in \mathcal{N}$.}
        \textbf{Initialization:} Set $\mathcal{N}_{\rm{empty}}=\mathcal{N}$ and $\mathcal{N}_{\rm{C}}=\varnothing$. Set the value for the size of UE subset $\mathcal{K}_r$, denoted by $S$, and record $\mathcal{G}^{\rm{cost}}_{\Pi}$ as the total MD radio resource costs as defined in $\mathcal{P}_{1}$ for each coalition formation $\Pi$\\
        \For{$k=1:K$}
        {
             \If{$\mathcal{N}_{\rm{empty}}\neq\varnothing$}
            {Randomly assign $k$-th UE to monopolize a coalition in $\mathcal{N}_{\rm{empty}}$}
            \Else{Find $n^* \in \mathcal{N}_{\rm{C}}$ satisfying Equation (\ref{EQ-9}) and constraints ${\rm {\tilde C}}1$, ${\rm C}2$\\
            Assign $k$-th UE to $\mathcal{U}_{n^*}$
            }
            Update UE assignment information for each $\mathcal{U}_n$ and record current coalition formation as $\Pi$.\\
            Update $\mathcal{N}_{\rm{empty}}$ and $\mathcal{N}_{\rm{C}}$
        }
        \While{\rm{the new ${\mathcal{K}_r}'$ has not been searched before}}
        {   
           \For{$s=1:S-1$}
           {
                \If{$\Pi_{\mathcal{K}_r,\xi_s}$ \rm{contains no \textit{forbidden coalition} and} $\mathcal{G}^{\rm{cost}}_{\Pi_{\mathcal{K}_r,\xi_s}}\leq\mathcal{G}^{\rm{cost}}_{\Pi}$}
                {
                    Set ${\mathcal{G}^{\rm{cost}}_{\Pi}}^{*}=\mathcal{G}^{\rm{cost}}_{\Pi_{\mathcal{K}_r,\xi_s}}$\\  
                    Record current matching result with respect to $\xi_s$
                }
           }
           Remove the examined $\mathcal{K}_r$ from the candidate list\\
           Construct a new UE subset ${\mathcal{K}_r}'$ \\
        }
    \end{scriptsize}
\end{algorithm}
\vspace{-0.3 cm}
\subsubsection{Computational Complexity} The overall complexity of the designed matching algorithm has been analyzed with respect to the total number of rotation operations. In \textbf{Algorithm \ref{alg1}}, the establishment of a feasible solution takes $K$ iterations to allocate each UE, the complexity is dominated by the rotation sequence process. Given the maximum number of iterations $N_{\rm iter}$, the overall complexity of \textbf{Algorithm \ref{alg1}} is bounded by \begin{small}$N_{\rm iter}\cdot(S-1)\cdot C^{S}_{K}\approx\mathcal{O}(N_{\rm iter}\cdot K^{S})$\end{small}, where $S \le 2$. {Specifically, the computational complexity can be controlled by selecting an appropriate value of $S$, which strikes a balance between the algorithm performance and the acceptable computational complexity. In the implementation of \textbf{Algorithm \ref{alg1}}, we recommend setting the value of $S$ as 3.}
\subsection{Stage II--Real-time Multi-dimensional Resource Allocation}
Upon deriving the UE coalition and its corresponding multiple access mode \begin{small}$\left({\mathcal{U}_n^{*}},{\alpha_n^{*}},{\beta_n^{*}}\right)$\end{small}, the original problem $\mathcal{P}$ is reduced to a joint subchannel and power allocation problem, which is given by\begin{small}
\begin{align}
&\mathcal{P}_2:\;\;\mathop{\max}\limits_{\left\{s_{n,m},p_k\right\}}  \left\{ { \sum\nolimits_{n \in \mathcal{N}} \sum\nolimits_{k \in \mathcal{U}_{n}} {{u_k}} } \right\},\;\;\text{s.t.:}\;{\rm{C3,C4,C5.}} \notag
\end{align}\end{small}However, subproblem $\mathcal{P}_{2}$ is non-concave, since that the achievable data rate, i.e. equation (\ref{data_rate}), in the objective function is a difference of concave functions. For this reason, subproblem $\mathcal{P}_{2}$ is difficult to solve efficiently for the global optimal solution.
\subsubsection{Subproblem Transformation}
Following the idea in our previous work \cite{Liu_TWC}, we restore to an efficient approximation method, referred to as successive convex approximation (SCA)~\cite{SCA}, and transform subproblem $\mathcal{P}_{2}$ into a concave optimization problem.
\begin{proposition}[\textbf{Transformation of Problem $\mathcal{P}_{2}$}]
 $\mathcal{P}_{2}$ can be approximated by the following concave maximization problem, which is given by\begin{small}
\begin{align}
&\tilde{\mathcal{P}}_{2}:\;\;\mathop{\max}\limits_{\left\{s_{n,m},z_{k,m}\right\}}  \left\{ { \sum\nolimits_{n \in \mathcal{N}} \sum\nolimits_{k \in \mathcal{U}_{n}} {\left({\bar r}_{k}/R_{\max} - w_k \cdot {\sum\nolimits_{m = 1}^{M} s_{n,m} \cdot g_{k,m}}\right)} } \right\} \notag\\
&\text{s.t.:}\;\;p_k=\sum\nolimits_{m=1}^{M} s_{n,m}\cdot e^{z_{k,m}}, \forall k \in {\mathcal{U}_n},n\in\mathcal{N},\nonumber\\
&\;\;\;\;\;\;\; {\bar r_k} = \sum\nolimits_{m = 1}^M {{s_{n,m}} \cdot {\frac{B}{M} \cdot \left[ {{a_{k,m}}{{\log }_2}\left( {{\gamma _{k,m}}} \right) + {b_{k,m}}} \right]} },\forall k \in {\mathcal{U}_n},n\in\mathcal{N},\nonumber \\
&\;\;\;\;\;\;\;{\rm{C3,C4}}: R_{\rm{th}} \le {{\bar r}_{k}} \le R_{\max }, \forall k \in \mathcal{K}, {\rm{C5,}}\nonumber
\end{align}
\end{small}where the transmit power of each UE $p_k$ is replaced by an equivalent form,
$p_k=\sum\nolimits_{m=1}^{M} s_{n,m}\cdot e^{z_{k,m}}$, \begin{small}$k \in \mathcal{U}_n$\end{small} and term ${\bar r_k}$ is the lower bound of the data rate of UE $k$ in equation (\ref{data_rate}), that is\begin{small}
\begin{equation} \label{SCA_formula}
    {\bar r_k} = \sum\nolimits_{m = 1}^M {{s_{n,m}} \cdot {\frac{B}{M} \cdot \left[ {{a_{k}}{{\log }_2}\left( {{\gamma _{k}}} \right) + {b_{k}}} \right]} }  \le {r_k}, k \in {\mathcal{U}_n}.
\end{equation}
\end{small}Herein, the constants $\left\{a_{k},b_{k}\right\}$ are chosen as specified bellow,\begin{small}
\begin{align}
    &a_{k} = \frac{{\gamma _k^{{\rm{th}}}}}{{1 + \gamma _k^{{\rm{th}}}}}, \tag{\ref{SCA_formula}a} \\
    &b_{k} = {\log _2}\left({1 + \gamma _k^{{\rm{th}}}}\right) - \frac{{\gamma _k^{{\rm{th}}}}}{{1 + \gamma _k^{{\rm{th}}}}}{\log _2}\left( {\gamma _k^{{\rm{th}}}} \right), \tag{\ref{SCA_formula}b}
\end{align}
\end{small}where ${\gamma_k^{{\rm{th}}}}=2^{M{R_{\rm th}}/B}-1$ is the lower bound of SINR for UE $k$. 
\end{proposition}
\begin{IEEEproof}
Firstly, we consider a relaxation of achievable data rate in equation (\ref{data_rate}) to avoid the structure of difference of concave functions. We make use of SCA to derive inequality (\ref{SCA_formula}). Furthermore, to strictly guarantee constraint C4, the lower bound of data rate, ${\bar r_k}$, is tight at ${\bar r_k}=R_k^{\max}$ when constants $\left\{a_{k},b_{k}\right\}$ are selected above. In the lower bound, ${\bar r_k}$, let us replace the transmit power of each UE, $p_k$, by a equivalent form $p_k=\sum\nolimits_{m=1}^{M} s_{n,m}\cdot e^{z_{k,m}}$, \begin{small}$k \in \mathcal{U}_n$\end{small}. Then, ${\bar r_k}$, is concave with respect to variables $\left\{s_{n,m},z_{k,m}\right\}$,\begin{small}\begin{equation}
    {{\bar r}_k} = \sum\nolimits_{m = 1}^M {{s_{n,m}} \cdot \frac{B}{M} \cdot \left[ {{a_k}{{\log }_2}\left( {{{\left| {{\boldsymbol{h}}_{k,m}^{\rm{H}}{{\boldsymbol{v}}_k}} \right|}^2}} \right) + {z_{k,m}} - {{\log }_2}\left( {{I_{k,m}}{\rm{ + }}{N_0}} \right) + {b_k}} \right]}, \notag
\end{equation}
\end{small}since ${\bar r_k}$ is composed by a sum of linear and concave terms (log-sum-exp is convex) \cite{Convex}. 
In the same way, $- w_k \cdot g_{k,m}$ is also a linear or concave function under all kinds of multiple access modes. Therefore, apply ${{\bar r}_k}$ to subproblem $\mathcal{P}_2$, we can obtain an standard concave maximization problem $\tilde{\mathcal{P}}_{2}$ in variables \begin{small}$\left\{ {{s_{n,m}},{z_{k,m}}\left| {\forall n \in {\mathcal{N}},k \in {{\mathcal{U}}_n},m \in \mathcal{M}} \right.} \right\}$\end{small}.
\end{IEEEproof}
\par
$\tilde{\mathcal{P}}_{2}$ is a concave maximization problem that can be solved by convex optimization approaches.
\subsubsection{Lagrange Dual Decomposition Method for Solving Subproblem $\tilde{\mathcal{P}}_{2}$}
Although subproblem $\tilde{\mathcal{P}}_{2}$ is concave, $\tilde{\mathcal{P}}_{2}$ is a mixed-integer optimization problem, which is hard to be solved. Therefore, to utilize the Lagrange dual decomposition method, we first form the
partial Lagrangian of $\tilde{\mathcal{P}}_{2}$ without considering constraint C3, which is given by\begin{small}
\begin{align}
    L({\boldsymbol \eta},{\boldsymbol \mu},{\nu},\left\{s_{n,m},z_{k,m}\right\})&=\sum\nolimits_{n \in \mathcal{N}} \sum\nolimits_{k \in {\mathcal U}_{n}}\left[{\bar r}_{k}/R_{\max} - w_k \cdot {\sum\nolimits_{m = 1}^{M} s_{n,m} \cdot g_{k,m}}\right] \nonumber\\ 
    &+\sum\nolimits_{n \in \mathcal{N}} \sum\nolimits_{k \in {\mathcal U}_{n}} \left[
    {\eta_k}\cdot\left({\bar r}_{k}-R_{\rm{th}}\right) - {\mu_k}\cdot\left({\bar r}_{k}-R_{\max}\right) \right] \nonumber\\
    &- \nu \cdot {\left(\sum\nolimits_{n \in \mathcal{N}}\sum\nolimits_{k \in {\mathcal U}_{n}}\sum\nolimits_{m=1}^{M}s_{n,m}\cdot e^{z_{k,m}} - P_{\max}\right)},
\end{align}%
\end{small}where $\boldsymbol{\eta} = \{\eta_k,\forall k\}$ is the Lagrange multiplier vector associated with the data rate constraints of each UE in inequality C4, and ${\nu}$ is the Lagrange multiplier related to the maximum transmission power of BS in constraint C6. 
\par
Then, the Lagrange dual function $J({\boldsymbol \eta},{\boldsymbol \mu},{\nu})$ is derived by solving the following problem,\begin{small}
\begin{equation}\label{J_fun}
    J\left({\boldsymbol \eta},{\boldsymbol \mu},{\nu}\right)=\mathop {\max }\limits_{\left\{s_{n,m},z_{k,m}\right\}} L({\boldsymbol \eta},{\boldsymbol \mu},{\nu},\left\{s_{n,m},z_{k,m}\right\}),\;{\rm subject\;to\; C3}.
\end{equation}
\end{small}Therefore, now the key point is to derive the value of Lagrange dual function $J({\boldsymbol \eta},{\boldsymbol \mu},{\nu})$ by solving problem (\ref{J_fun}). The basic track of solving problem (\ref{J_fun}) is as follows: Firstly, without considering constraints in problem (\ref{J_fun}) and setting $s_{n,m}=1, \forall n,m$, then problem (\ref{J_fun}) can be decomposed into $M^2$ individual problems, which is\begin{small}
\begin{align}\label{individual_problem}
   &{J_{n,m}}\left( {{\boldsymbol{\eta }},{\boldsymbol{\mu }},\nu } \right) = \mathop {\max }\limits_{\left\{ {{z_{k,m}}\left| {k \in {{\mathcal U}_n}} \right.} \right\}} \sum\limits_{k \in {{\cal U}_n}} {\left[ {\left( {1/{R_{\max }} + {\eta _k} - {\mu _k}} \right) \cdot \frac{B}{M} \cdot {a_k}{{\log }_2}\left( {{\gamma _k}} \right) - \nu  \cdot {e^{{z_{k,m}}}} - {w_k} \cdot g_{k,m}} \right]}, \notag \\
   & \forall n \in \mathcal{N}, 1 \le m \le M.\tag{\ref{J_fun}a}\end{align}\end{small}If the $m$-th SC is assigned to the $n$-th UE coalition, then the optimal power allocation of UEs in coalition $n$ can be obtained by solving problem (\ref{individual_problem}). Then, considering constraint C3, the optimal solution to problem (\ref{J_fun}) and dual function value $J({\boldsymbol \eta},{\boldsymbol \mu},{\nu})$ can be obtained.\par
\begin{proposition}\label{lemma_1}
Given the multiple access mode indicators $\{\alpha_n^*, \beta_n^*\}$ for coalition \begin{small}$\mathcal{U}_n$\end{small}, the optimal solution \begin{small}$\{ {{z_{k,m}^{*}}\left| {k \in {{\mathcal U}_n}} \right.} \}$\end{small} to problem (\ref{J_fun}a) is represented by,
\begin{itemize}
    \item if $\alpha_n^* = 1$ and $\beta_n^* = 0$ (i.e., power-domain NOMA mode is chosen for \begin{small}$\mathcal{U}_n$\end{small}), then
    \begin{small}
        \begin{empheq}[left={{{\rm e}^{{z_{k,m}^ * }}}=\empheqlbrace}]{alignat=2}
       & {\min \left\{\frac{{\gamma^{{\max}}}\cdot (N_0)}{| \boldsymbol{h}^{\rm H}_{k,m}\boldsymbol{v}_{k}|^{2}}, \frac{D_k}{\nu}\right\}\;{\rm if\;}k{\rm \; is\;near\;UE\;in\;the\;power\;domain\;NOMA\;pair},\tag{13a}} \\
       & {\min \left\{\frac{{\gamma^{{\max}}}\cdot (I_{k,m}^{\rm PD}+N_0)}{| \boldsymbol{h}^{\rm H}_{k,m}\boldsymbol{v}_{k}|^{2}}, \frac{D_k}{\nu}\right\},\;{\rm if\;}k{\rm \; is\;far\;UE},\tag{13b}}
        \end{empheq}
    \end{small}where ${\gamma^{{\max}}}=2^{M{R_{\max}}/B}-1$.
    \item if $\alpha_n^* = 0$ and $\beta_n^* = 0\;{\rm or}\;1$ (i.e., OMA or spatial-domain NOMA mode is chosen for \begin{small}$\mathcal{U}_n$\end{small}), then we have the following fixed-point equation,
    \begin{small}
    \begin{equation}\label{fixed_point_eq}
    {\rm e}^{{z_{k,m}^ * }} = \frac{{{D_k}}}{{\nu  + \sum\limits_{i \in {{\mathcal{U}}_n}\backslash \left\{ k \right\}} {{G_{i,k,m}} \cdot {\gamma _{i,m}} \cdot {\rm e}^{-{z_{i,m}^ * }}} }}, \tag{14}
    \end{equation}
    \end{small}where constants $D_k$ and $G_{i,k,m}$ are defined to simplify the notation,\begin{small}
    \begin{align}
    D_{k} & = \frac{1}{{\ln 2}} \cdot \left(\frac{1}{R_{\max}}+{\eta}_k-{\mu}_k\right)\cdot{\frac{B}{M} \cdot {a_k}}, \tag{\ref{fixed_point_eq}a}\\
    G_{i,k,m} & = \frac{1}{{\ln 2}} \cdot \left(\frac{1}{R_{\max}}+{\eta}_i-{\mu}_i\right)\cdot{\frac{B}{M} \cdot {a_i}} \cdot \frac{{{{\left| {h_{i,m}^{\rm{H}}{v_k}} \right|}^2}}}{{{{\left| {h_{i,m}^{\rm{H}}{v_i}} \right|}^2}}}. \tag{\ref{fixed_point_eq}b} 
    \end{align}\end{small}
    \item if $\alpha_n^* = 1$ and $\beta_n^* = 1$ (i.e., hybrid NOMA mode is chosen for \begin{small}$\mathcal{U}_n$\end{small}), then
    \begin{small}
    \begin{align}\label{fixed_point_eq_2}
    {\rm e}^{{z_{k,m}^ * }} = \frac{{{D_k}+{E_k}}}{{\nu  + \sum\limits_{i \in {{\mathcal{U}}_n}\backslash \left\{ k \right\}} ({{F_{i,k,m}} \cdot {\gamma _{k,m}^{\rm sic}} } + {{G_{i,k,m}} \cdot {\gamma _{i,m}} })\cdot {\rm e}^{-{z_{i,m}^ * }} } + 
    \sum\limits_{(i,i') \in {{\mathcal{U}}_n}\backslash \left\{ k \right\}} V_{i,i',k,m} \cdot {\gamma _{i,m}^{\rm sic}}
    \cdot {\rm e}^{-{z_{i',m}^ * }}
    }, \tag{15}
    \end{align}
    \end{small}where constants $E_{k}$, $F_{i,k,m}$, $G_{i,k,m}$ and $V_{i,i',k,m}$ are defined as\\
    \begin{small}
    \begin{empheq}[left={E_{k}=\empheqlbrace}]{alignat=2}
       & {{\frac{\rho_{1} w_k}{{\ln 10}}},\;{\rm if\;}k{\rm \; is\;the\;far\;UE}{\rm\;in\;a\;power\;domain\;NOMA\;pair}},
        \quad\quad\quad\quad\quad\quad\quad\quad\quad\quad	\tag{\ref{fixed_point_eq_2}a}\\
       & {{0},\;{\rm{otherwise,}}\tag{\ref{fixed_point_eq_2}b}} 
    \end{empheq}
    \begin{empheq}[left={F_{i,k,m}=\empheqlbrace}]{alignat=2}
        & \frac{\rho_{1} w_k}{{\ln 10}} \cdot \frac{{{{| {{\boldsymbol h}_{k,m}^{\rm{H}}{{\boldsymbol v}_k}} |}^2}}}{{{{| {{\boldsymbol h}_{k,m}^{\rm{H}}{{\boldsymbol v}_i}}|}^2}}},\;
        {{\rm if\;}k{\rm \; is\;near\;UE\;and\;}i{\rm\;is\;far\;UE\;in\;a\;power\;domain\;NOMA\;pair}}, \tag{\ref{fixed_point_eq_2}c} \\
        & {{0},\;{\rm{otherwise,}}\tag{\ref{fixed_point_eq_2}d}} 
    \end{empheq}
    \begin{empheq}[left={G_{i,k,m}=\empheqlbrace}]{alignat=2}
        & 0,\;{{\rm if\;}k{\rm \; is\;far\;UE\;and\;}i{\rm\;is\;near\;UE\;in\;a\;power\;domain\;NOMA\;pair}},
        \quad\quad\quad\quad\quad\quad\quad 
        \tag{\ref{fixed_point_eq_2}e}\\
        & {{\frac{1}{{\ln 2}} \cdot \left(1/{R_{\max}}+{\eta}_i-{\mu}_i\right)\cdot{\frac{B}{M} \cdot {a_i}} \cdot \frac{{{{| {{\boldsymbol h}_{i,m}^{\rm{H}}{{\boldsymbol v}_k}}|}^2}}}{{{{| {{\boldsymbol h}_{i,m}^{\rm{H}}{{\boldsymbol v}_i}}|}^2}}}},\;{\rm{otherwise,}}\tag{\ref{fixed_point_eq_2}f}} 
    \end{empheq}
    \begin{empheq}[left={V_{i,i',k,m}=\empheqlbrace}]{alignat=2}
        & \frac{\rho_{1} w_k}{{\ln 10}} \cdot \frac{{{{| {{\boldsymbol h}_{i,m}^{\rm{H}}{{\boldsymbol v}_k}} |}^2}}}{{{{| {{\boldsymbol h}_{i,m}^{\rm{H}}{{\boldsymbol v}_{i'}}}|}^2}}},\;{{\rm if\;}(i,i'){\rm \;is\;near\;and\;}{\rm far\;UE\;in\;a\;power\;domain\;NOMA\;pair}},
        \quad\;\;\;\; 
        \tag{\ref{fixed_point_eq_2}g}\\
        & {0,\;{\rm{otherwise.}}\tag{\ref{fixed_point_eq_2}h}} 
    \end{empheq}
    \end{small}
\end{itemize}
\begin{IEEEproof}
The temporary optimal solution $z_{k,m}^{*}$ can be calculated as follows by setting the gradient of the objective function in (\ref{J_fun}a) to be zero. Hence, we can obtain the fixed-point equation (\ref{fixed_point_eq}). Then, optimal allocation power can be updated iteratively with the fixed-point equation (\ref{fixed_point_eq}), in which convergence is easily guaranteed since the right-hand side of (\ref{fixed_point_eq}) is a standard interference function \cite{JSAC_1995}.
\end{IEEEproof}
\end{proposition}\par
After obtaining the optimal solution to individual problems (\ref{J_fun}a) based on \textbf{Proposition \ref{lemma_1}},  we can derive the value of Lagrange dual function $J({\boldsymbol \eta},{\boldsymbol \mu},{\nu})$,
\begin{proposition}[\textbf{Optimal Solution to Problem (\ref{J_fun})}] \label{J_fun_solution}
 Firstly, the optimal SC allocation for UE coalitions with fixed $\left( {{\boldsymbol{\eta }},{\boldsymbol{\mu }},\nu } \right)$ can be decided based on the following criterion,\begin{small}
 \begin{equation}
     s_{n,m}^{*} = \argmax \nolimits_{\left\{n' \in \mathcal{N}, 1 \le m' \le M\right\}} \left\{\sum\nolimits_{n',m'} {J_{n',m'}}\left( {{\boldsymbol{\eta }},{\boldsymbol{\mu }},\nu } \right)\right\},\;{\rm subject\;to\; C3}, \notag
 \end{equation}
 \end{small}which fits a maximum weight matching for bipartite graphs and can be solved by the Hungarian algorithm. Then, the corresponding power allocation of UEs in each coalition can be given by\begin{small}\begin{equation}
     p_k^{*} = {\sum\nolimits_{m = 1}^M {s_{n,m}^ *  \cdot \exp \left( {z_{k,m}^ * } \right)} }. \notag
 \end{equation}\end{small}\end{proposition}
\par
In this way, the corresponding dual problem to the original problem $\tilde{\mathcal{P}}_{2}$ is\begin{small}
\begin{align}\label{dual_problem}
	& \quad \mathop {\min }\limits_{{\boldsymbol{\eta }},{\boldsymbol{\mu }},\nu} \;\left\{ {{{J}}\left( {{\boldsymbol{\eta }},{\boldsymbol{\mu }},\nu} \right)} \right\},\;{\rm{subject\;to\;}}{\boldsymbol{\eta }} \ge {\boldsymbol{0}},{\boldsymbol{\mu }} \ge {\boldsymbol{0}},{\nu} \ge {0}. \tag{16}
\end{align}
\end{small}By solving problem (\ref{dual_problem}), we can obtain the optimal value for $\tilde{\mathcal{P}}_{2}$ because of the strong duality. Since the dual function $J\left({\boldsymbol \eta},{\boldsymbol \mu},{\nu}\right)$ is the point-wise infimum of a set of affine functions of Lagrange multipliers, it is a typical concave function, which can be solved by standard sub-gradient method.
\par
Since the corresponding dual problem (\ref{dual_problem}) to the original problem $\tilde{\mathcal{P}}_{2}$ is convex, it can be solved by using sub-gradient iteration method \cite{Convex}. For readers' convenience, the processes of solving $\tilde{\mathcal{P}}_{2}$ are presented in \textbf{Algorithm \ref{alg2}}. Finally, we solve problem $\tilde{\mathcal{P}}_{2}$ and obtain the optimal UE coalition formation, multiple access mode selection, subchannel, and power allocation scheme, \begin{small}
$\Omega^{*}=\left\{(\mathcal{U}_n^{*}, \alpha_n^{*}, \beta_n^{*}), s_{n,m}^{*}, p_{k}^{*}\right\}$\end{small}.
\subsubsection{Computational Complexity} 
In conclusion, for solving Problem $\mathcal{P}_2$, the computation complexity of each iteration in \textbf{Algorithm \ref{alg2}} is divided into three parts: a) using \textbf{Lemma \ref{lemma_1}} to solve problem (\ref{J_fun}a), which involves $M^2$ parallel streams; b) using \textbf{Proposition \ref{J_fun_solution}} to perform SC and power allocation under given Lagrange multipliers $\boldsymbol{\eta}$, $\boldsymbol{\mu}$ and $\nu$; c) the update of Lagrange multipliers $\boldsymbol{\eta}$, $\boldsymbol{\mu}$ and $\nu$.  The main complexity of \textbf{Algorithm 2} is  induced by part a) and b), which makes the computation complexity of \textbf{Algorithm \ref{alg2}} is upper bounded by \begin{small}$\mathcal{O}\left(M^2 \cdot L_{\max} + M^3\right)$.\end{small}
\begin{algorithm}
    \setstretch{1} 
    \caption{Joint SC and Power Allocation to UE coalitions}
    \label{alg2}
    \begin{scriptsize}
    	\SetKwInOut{Input}{Input}\SetKwInOut{Output}{Output}
        \Input{UE Coalition, $\left\{(\mathcal{U}_n^{*}, \alpha_n^{*}, \beta_n^{*}), \forall n \right\}$ and channel conditions of each UE, $\left\{\boldsymbol{h}_{k,m}, \forall k,m\right\}$} 
        
        \Output{SC allocation of UE coalition $\{ {{s_{n,m}^{*}}, {\forall n,m}}\}$, and power allocation of each UE $\{p_{k}^{*}, \forall k\}$.}
        \textbf{Initialization:} Set iteration index $t = 0$. Choose ${\nu}_{\min }^{(0)}$ and  ${\nu}_{\max }^{(0)}$ and set ${\nu}^{(0)}=0$ and tolerable error $\epsilon$. Set Lagrange multiplier vectors $\boldsymbol{\eta}=\boldsymbol{\mu}=\boldsymbol{0}$. \\
        \While{$| {\nu_{\max }^{(t)} - \nu_{\min }^{(t)}}| > \epsilon$}
        {   \While{$| {J^{(t)} -J^{(t-1)}}| > \epsilon$ {\rm \textbf{and}} $t \ge 1$}
            {By using \textbf{Proposition \ref{lemma_1}} and \textbf{Proposition \ref{J_fun_solution}}, solve Problem (\ref{J_fun}) obtain the optimal solution $\{s_{n,m}^*,p_{k}^*\}$ with fixed value of Lagrange multipliers $\{\boldsymbol{\eta},\boldsymbol{\mu},\nu^{(t)}\}$.\\
            Set $J^{(t)} \leftarrow  J\left({\boldsymbol \eta},{\boldsymbol \mu},{\nu}^{(t)}\right)$\\
            Update $\boldsymbol{\eta}$, $\boldsymbol{\mu}$ using the subgradient method,\\
            $\quad\quad{{\eta}_k}  \leftarrow {{\eta}_k} - \Delta \cdot \left(r_k^{*} - R{\rm th}\right)\;,\forall k \in {\mathcal{U}_n},n\in\mathcal{N}$, \\
            $\quad\quad{{\mu}_k}  \leftarrow {{\mu}_k} - \Delta \cdot \left(R{\max} - r_k^{*}\right),\;\forall k \in {\mathcal{U}n},n\in\mathcal{N}$, \\
            where $\Delta$ is the stepsize.
           }
           \If{$\sum\nolimits_{k \in \mathcal{K}} p_{k}^* \le P_{\max}$ {\rm \textbf{and}} $t == 0$}
            {\textbf{Break}}
            Update $\nu$ by ${\nu}^{(t+1)} \leftarrow [\nu_{\max }^{(t)}+\nu_{\min }^{(t)}]/2$.\\
            \If{$\sum\nolimits_{k \in \mathcal{K}}  p_{k}^* > P_{\max}$}
            {${\nu}_{\min}^{(t+1)} \leftarrow {\nu}^{(t+1)}$}
            \Else{${\nu}_{\max}^{(t+1)} \leftarrow {\nu}^{(t+1)}$}
        }
    \end{scriptsize}
\end{algorithm}
\section{Simulation Results and Analysis}\label{Simulation}
\subsection{Simulation Setup}
In this section, a system-level simulation platform is implemented. Simulation results are provided in this section to evaluate the proposed multi-dimensional multiple access scheme. For simulation parameters and deployments, we consider a single cell and set the maximum transmit power of BS as 33 dBm, the bandwidth of SC as 2 MHz, the number of subchannels as 12, noise variance as -140 dBm/Hz and the maximum number of UEs of one coalition is 3. Moreover, the default number of UEs is 25. Detailed simulation parameters are summarized in Table~I. 
\begin{table}[!h]
	\centering
	\newcommand{\tabincell}[2]{\begin{tabular}{@{}#1@{}}#2\end{tabular}}
	\scriptsize
	\renewcommand{\arraystretch}{1.0}
	\caption{Default Parameter settings.}
		\label{parameters}
		\begin{tabular}{|l|l|}
			\hline
			\textbf{Parameter} & \textbf{Assumption}\\
			\hline
			Bandwidth/Number of SCs & 20 MHz/12\\
			\hline
			Rician fading factor $\kappa$ & 9\\
			\hline
			Distance-dependent Path loss (dB) & $PL(R) = 128.1 + 37.6 \times \lg(R),\;R\;\rm{in}\;km$\\
			\hline
			Macro cell radius/Cellular UE Distribution & $250\;m$/Uniform in the Marco cell\\
			\hline
			Maximum transmit power of BS & 33 dBm \\
			\hline
			Number of UEs in the Marco cell & 25 \\ 
			\hline
			Number of antennas & 64 TX antenna for BS and 1 RX antenna for UE \\
			\hline
			Pathloss model & WINNER+ B1 LOS \\ 
			\hline
			AWGN power & -140 dBm/Hz \\
			\hline
			Number of beamspaces & 10 \\
			\hline
			Coefficients related to the utilization cost & $\rho_{0} = 0.5$, $\rho_{1} = 0.24$, $\rho_{2} = 0.5$, and $w_k = 0.2$\\
			\hline
		\end{tabular}
	\end{table}
\par
To further evaluate the performance of proposed MDMA scheme, This paper compares the performance of our scheme with the following two baseline schemes:
\begin{itemize}
    \item \textit{MIMO-NOMA:} This scheme is realized by work \cite{MIMO-NOMA}, in which multiple UEs in the same beamspace are served by power-domain NOMA. Compared to our proposed scheme, the difference lies in no spatial-multiplexing among UEs that are located in different beamspaces.
    \item \textit{RSMA:} According to the RSMA scheme proposed by \cite{mao2018rate} and \cite{9461768}, in one SC, the message intended for one UE is split into the common part and private part: The common parts of all UEs in the same SC are combined into the common message and broadcast to UEs in the same SC; The private part containing the remaining part of the message is encoded into the private stream and transmitted by SDMA. Specifically, to make the scheme match our framework, we assign UEs to each SC following the paradigm of our previous work \cite{Wudan}.
\end{itemize}
\subsection{Experiment Result Analysis}
\begin{figure}[h]
    \vspace{-0.5cm}
    \centering
    \includegraphics[width=0.65\textwidth]{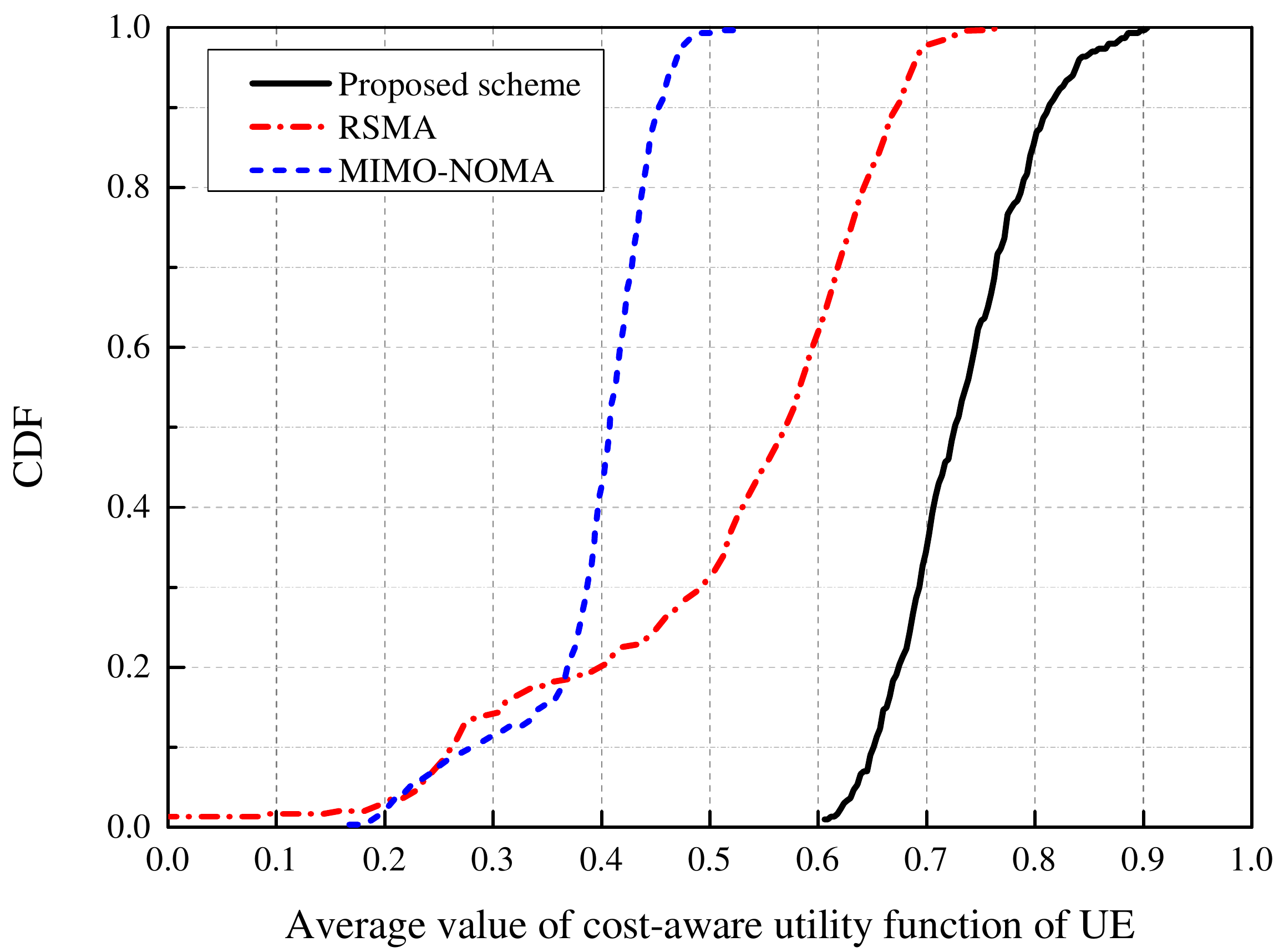}
    \caption{CDF of UE's cost-aware utility function defined in (\ref{U_f}), which can reflect the cost-efficiency of service provisioning. The higher value of the utility function means that the multiple access scheme can achieve the satisfied QoS performance with less utilization cost of UE. As we can see, the proposed MDMA scheme significantly outperforms the baseline schemes.}\centering
    \label{3}
\end{figure}
Fig. \ref{3} using the cumulative distribution function (CDF) provides an overview of performance gain achieved by three different multiple access schemes under the  deployment of 25 UEs. According to the obtained statistical information, the averaged utility curves for RSMA and MIMO-NOMA span from a minimal value of 0.22 to a ceiling of 0.64, and the latter from 0.23 to 0.54, respectively. As a contrast, the proposed MDMA scheme significantly outperforms the formers with distinguished numerical gains, which the increased upper-bound can prove.
\par
\begin{figure}[h]
    \centering
    \includegraphics[width=0.65\textwidth]{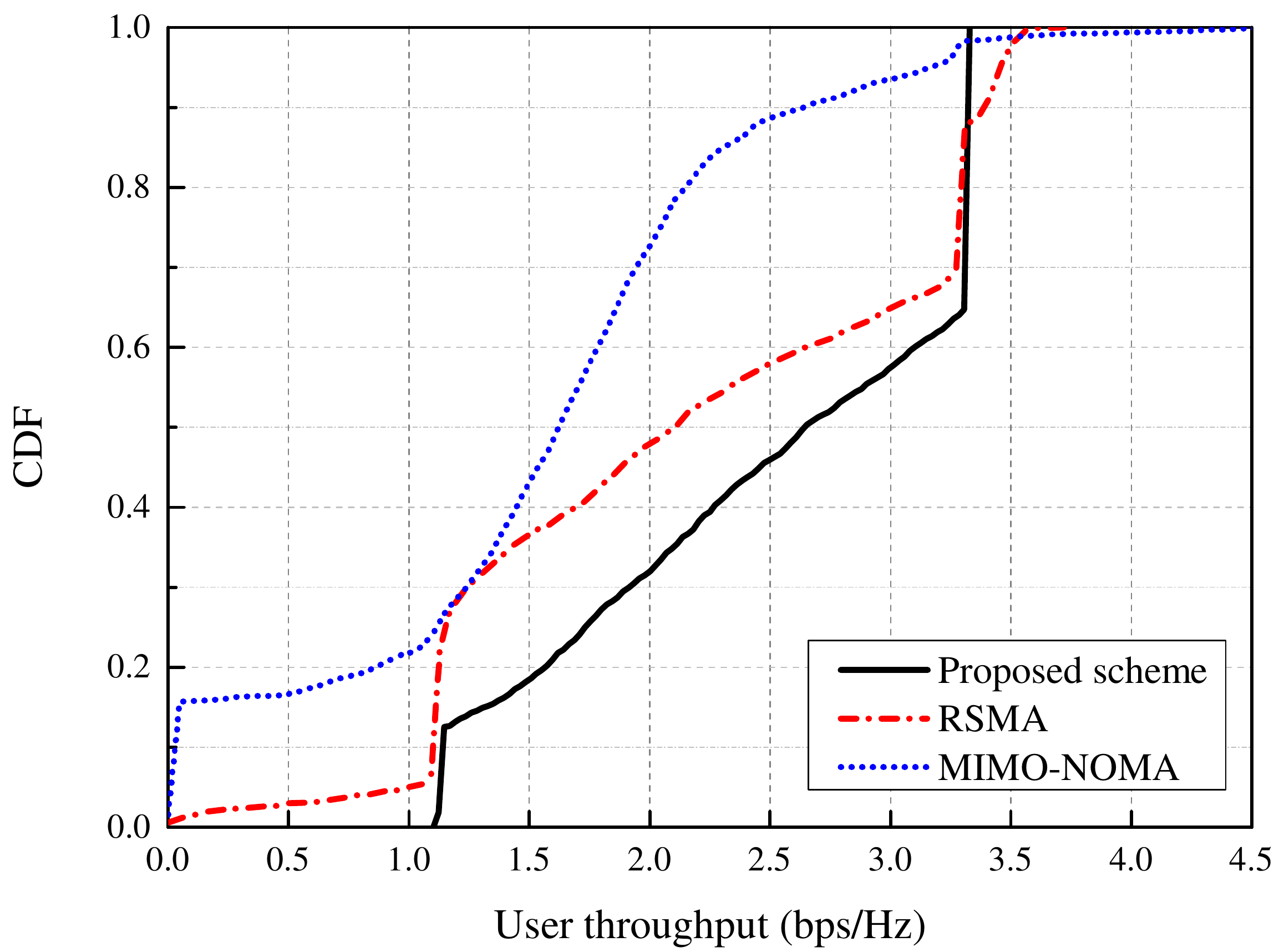}
    \caption{CDF of UE's downlink throughput. In this paper, the QoS performance is measured by the throughput of UE. The proposed MDMA scheme has better QoS performance than baseline schemes, which means our proposed scheme can better exploiting resource dimensions in frequency, spatial and power domain to fulfill UE's service demands according to the multi-dimensional radio resource availability.}\centering
    \label{data_rate}
    \vspace{-0.3cm}
\end{figure}
Fig.\ref{data_rate} shows the CDF of downlink throughput (in bps/Hz) experienced by UE. Firstly, the proposed MDMA scheme has achieved the best results with a relatively small variance in communication quality. Since the proposed MDMA can strictly guarantee the data rate of UE is between the minimum required data rate ($R_{\rm{th}}$) and the ideal maximum data rate ($R_{\max}$), the curve of MDMA has sudden rises at around 1.15 and 3.3 bps/Hz. Though RSMA and MIMO-NOMA are both well-established multiple-access schemes that have demonstrated great advantages in certain resource dimensions, our proposed MDMA is proved to be capable of reaping multiple dimensional multiplexing benefits according to varying network situations. In contrast, RSMA and MIMO-NOMA use fixed multiple access mode for UEs under all kinds of situations.
\par
\begin{figure}[!h]
\vspace{-0.5cm}
	\centering
	\subfigure[CDF of UE's utilization cost]{
		\begin{minipage}{11cm}
			\centering
			\includegraphics[width=11cm]{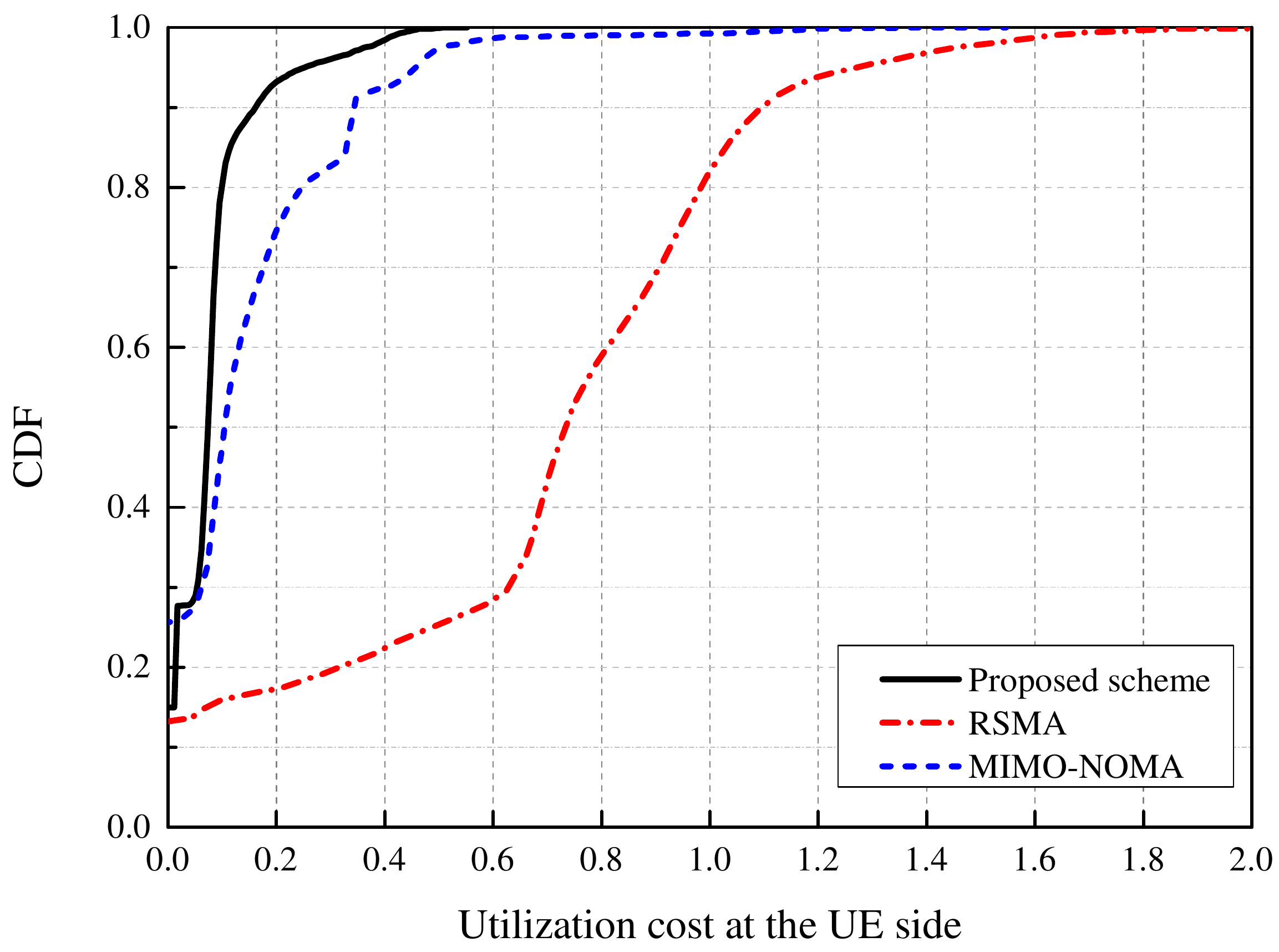}
		\end{minipage}
	}
	
	\subfigure[Scatter plot of utilization cost in different domains]{
		\begin{minipage}{11cm}
			\centering
			\includegraphics[width=11cm]{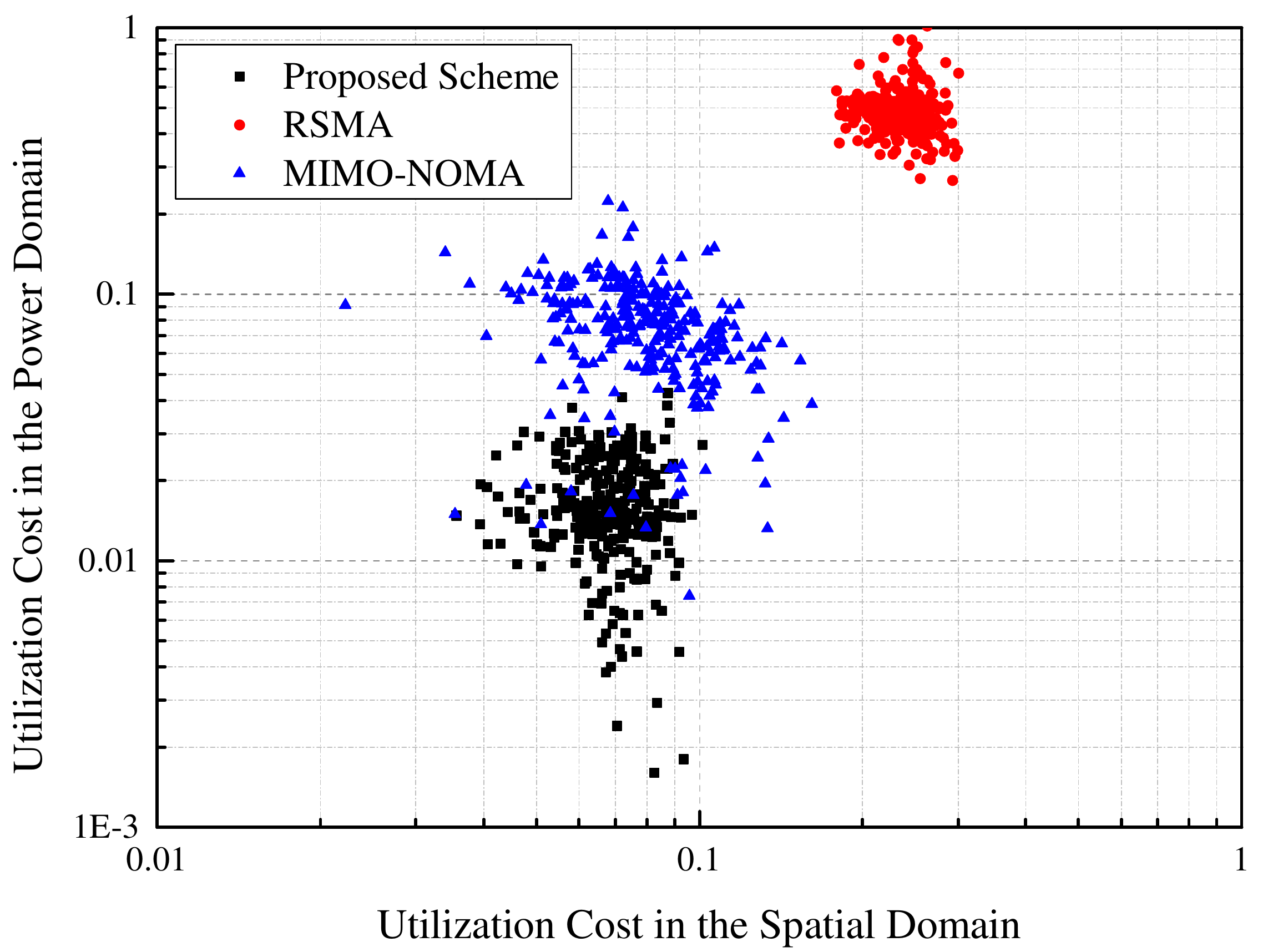}
		\end{minipage}
	}
	\caption{The performance of UE's resource utilization cost defined in (\ref{cost_formula}), which is summation of the processing cost of using power-domain and spatial-domain radio resources at the UE side under different multiple access modes, network conditions, and hardware capabilities. Our proposed scheme has lower utilization costs than baseline schemes, which use the fixed multiple access modes and resource allocation strategy to deal with UE-specific resource constraints and varying network conditions.}
	\label{4}
	\vspace{-0.5cm}
	\end{figure}
Fig.\ref{4}-(a) evaluates the multi-dimensional resource utilization costs under different multiple access schemes. Unlike RSMA and MIMO-NOMA with the medium cost value up to 0.18 and 0.63, the utilization cost under our proposed scheme remains lower values with an approximate average of 0.13. As discussed in Section \ref{coalition}, we designed a UE coalition strategy to mitigate non-orthogonal interference resulted from the multi-dimensional resource sharing and UE superposition, the resource sharing conflicts are largely reduced within each UE coalition. Hence, the proposed MDMA can adjust beneficial multiple access mode for each UE coalition flexibly.   
\par
Meanwhile, Fig.\ref{4}-(b) shows a scatter plot of the average utilization cost of UE in different domains during each simulation drop. Specifically, RSMA and MIMO-NOMA schemes are less effective in non-orthogonal interference mitigation process. For MIMO-NOMA method, UEs in each cluster experience aligned channel directions and a large disparity in channel strengths, whereas UEs in different clusters experience orthogonal channels. In consequence, some UEs located at disadvantageous cell coverage areas may suffer from severe interference. RSMA has a higher utilization cost than the other two schemes due to the following three reasons: i). UE needs to perform SIC to decode its private message, which will increase the cost in the power-domain especially for UEs with bad channel conditions; ii). The performance of RSMA is also relied on the precoder design for broadcasting the common message. However, the precoder design is an intractable problem; iii). The UE-SC (subcarrier) matching for multi-carrier RSMA is still an open issue, since different UE combinations for sharing one SC may induce a tremendous performance difference. 
\par
\begin{figure}[h]
    \vspace{-0.2cm}
    \centering
    \includegraphics[width=0.7\textwidth]{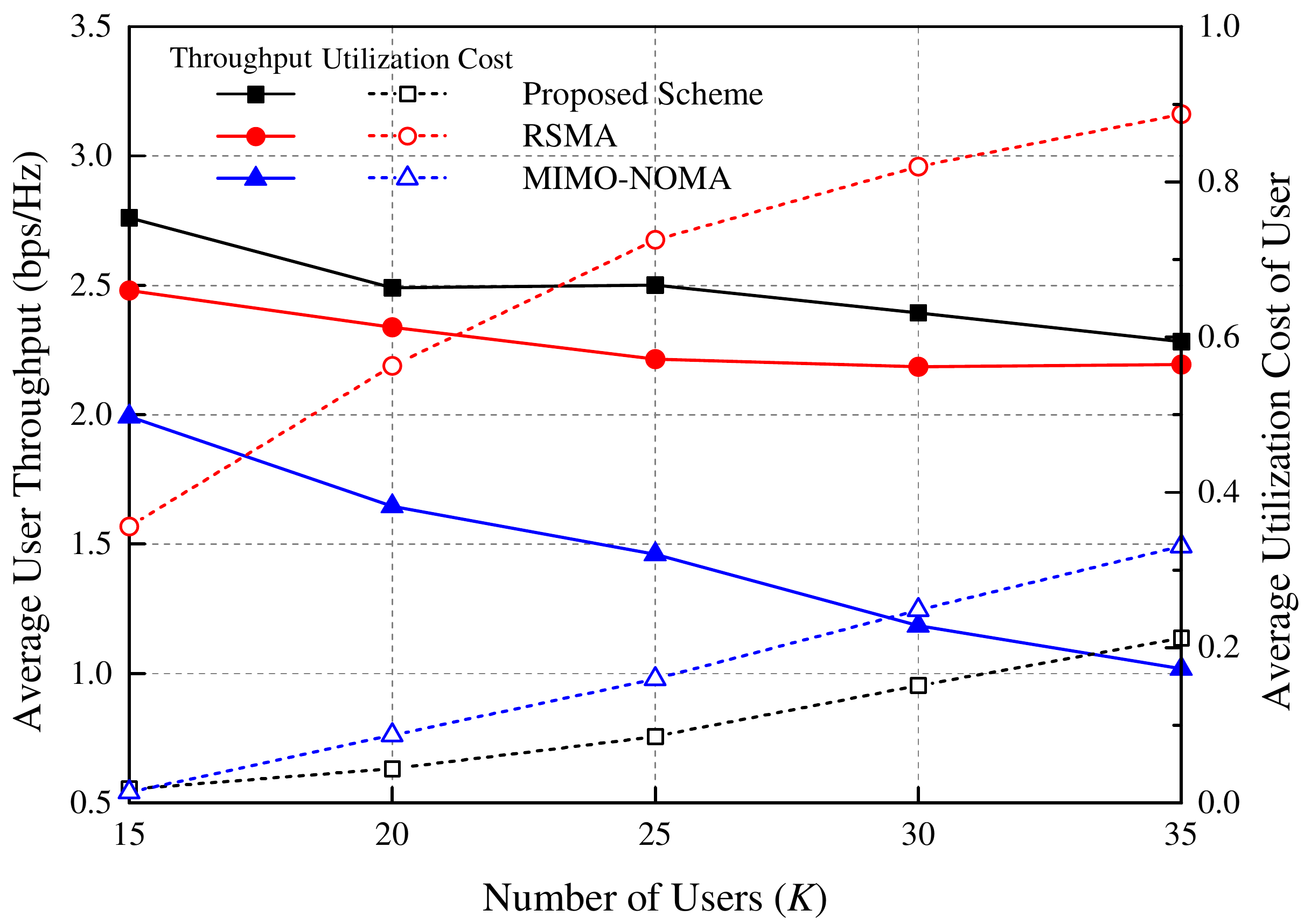}
    \caption{Average UE's downlink throughput of UE and utilization cost of radio resources under different multiple access schemes with respect to the number of served users. The results show our proposed scheme has the lowest utilization cost and highest throughput (QoS performance), and its "cost-gain" performance is robust to varying network traffic load.}\centering
    \label{5}
    \vspace{-0.2cm}
\end{figure}
Fig.~\ref{5} can be seen as a "cost-gain" comparison of several multiple access schemes under different user deployment scenarios, ranging from underloaded network situation to overloaded (reflected by the number of UEs). Compared with the SDMA and MIMO-NOMA, our proposed scheme has the lowest utilization cost and highest throughput, and its "cost-gain" performance is robust to varying network load and situation-dependent resource conditions. The reason is that RSMA and MIMO-NOMA use fixed multiple access mode for UEs under all scenarios, which will induce insufficient usage of the available multi-dimensional resource. It also shows that it is necessary to intelligently adapt multiple access modes of UE according to each user device's specific demands and specific resource situations. Overall, the result conforms to our expectations of the proposed MDMA scheme.
\section{Conclusions}\label{conclusion}
This paper has proposed a multi-dimensional multiple access (MDMA) scheme to meet the specific QoS demands and resource situations of each user in a wireless network cost-effectively. In our proposed scheme, each UE's QoS performance and resource constraints are jointly considered in the UE's cost-aware utility function. To maximize the summation of all UEs' utility functions, coexisting UEs with diverse QoS heterogeneity and resource diversity can be served efficiently. The formulated problem are separately in two stages, that is, cost-aware multiple access mode selection and multi-dimensional radio resource allocation. Based on the two-sided many-to-one matching theory, the user-specific multiple access mode is chosen to fully utilize the available multi-dimensional resources with acceptable utilization costs for users. Then, the multi-dimensional radio resource allocation problem is solved with moderate complexity based on convex optimization theory. Simulation results show that the proposed scheme is more effective than state-of-the-art schemes and outperforms them in both the users' QoS performance and multi-dimensional radio resource utilization cost.        
%
\bibliographystyle{IEEEtran}
\bibliography{IEEEabrv,Ref}

\begin{thebibliography}{10}
\providecommand{\url}[1]{#1}
\csname url@samestyle\endcsname
\providecommand{\newblock}{\relax}
\providecommand{\bibinfo}[2]{#2}
\providecommand{\BIBentrySTDinterwordspacing}{\spaceskip=0pt\relax}
\providecommand{\BIBentryALTinterwordstretchfactor}{4}
\providecommand{\BIBentryALTinterwordspacing}{\spaceskip=\fontdimen2\font plus
\BIBentryALTinterwordstretchfactor\fontdimen3\font minus
  \fontdimen4\font\relax}
\providecommand{\BIBforeignlanguage}[2]{{%
\expandafter\ifx\csname l@#1\endcsname\relax
\typeout{** WARNING: IEEEtran.bst: No hyphenation pattern has been}%
\typeout{** loaded for the language `#1'. Using the pattern for}%
\typeout{** the default language instead.}%
\else
\language=\csname l@#1\endcsname
\fi
#2}}
\providecommand{\BIBdecl}{\relax}
\BIBdecl

\bibitem{Kenneth}
``Global mobile data traffic from 2017 to 2022,'' J. Clement, Tech. Rep., 2019.

\bibitem{6g_cases}
M.~{Giordani}, M.~{Polese}, M.~{Mezzavilla}, S.~{Rangan}, and M.~{Zorzi},
  ``{Toward {6G} Networks: Use Cases and Technologies},'' \emph{IEEE Commun.
  Mag.}, vol.~58, no.~3, pp. 55--61, Mar. 2020.

\bibitem{you2021towards}
X.~You, C.~Wang, J.~Huang \emph{et~al.}, ``{Towards 6G wireless communication
  networks: Vision, enabling technologies, and new paradigm shifts},''
  \emph{Science China Information Sciences}, vol.~64, no.~1, pp. 1--74, Jan.
  2021.

\bibitem{Liu2}
Y.~Liu, X.~Wang, and J.~Mei, ``{Hybrid Multiple Access and Service-Oriented
  Resource Allocation for Heterogeneous {QoS} Provisioning in Machine Type
  Communications},'' \emph{Journal of Communications and Information Networks},
  vol.~5, no.~2, pp. 225--236, Jun. 2020.

\bibitem{AI-NS}
X.~{Shen}, J.~{Gao}, W.~{Wu}, K.~{Lyu}, M.~{Li}, W.~{Zhuang}, X.~{Li}, and
  J.~{Rao}, ``{{AI}-Assisted Network-Slicing Based Next-Generation Wireless
  Networks},'' \emph{IEEE Open Journal of Vehicular Technology (OJVT)}, vol.~1,
  pp. 45--66, Jan. 2020.

\bibitem{Ruitao}
R.~Chen and X.~Wang, ``{Maximization of Value of Service for Mobile
  Collaborative Computing through Situation Aware Task Offloading},''
  \emph{IEEE Trans. Mob. Comput.}, pp. 1--1, Jun. 2021.

\bibitem{liu2021application}
Y.~{Liu}, W.~{Yi}, Z.~{Ding}, X.~{Liu}, O.~{Dobre}, and N.~{Al-Dhahir},
  ``Application of {NOMA} in {6G} networks: Future vision and research
  opportunities for next generation multiple access,'' \emph{arXiv preprint
  arXiv:2103.02334}, Mar. 2021.

\bibitem{liu2021evolution}
Y.~{Liu}, S.~{Zhang}, X.~{Mu}, Z.~{Ding}, R.~{Schober}, N.~{Al-Dhahir},
  E.~{Hossain}, and X.~{Shen}, ``Evolution of {NOMA} toward next generation
  multiple access (ngma),'' \emph{arXiv preprint arXiv:2108.04561}, Aug. 2021.

\bibitem{Dai}
R.~Jiao and L.~Dai, ``{On the Max-Min Fairness of Beamspace {MIMO-NOMA}},''
  \emph{IEEE Trans. Signal Process.}, vol.~68, pp. 4919--4932, Aug. 2020.

\bibitem{mao2018rate}
Y.~Mao, B.~Clerckx, and V.~O. Li, ``{Rate-Splitting Multiple Access for
  Downlink Communication Systems: Bridging, Generalizing, and Outperforming
  SDMA and NOMA},'' \emph{EURASIP journal on wireless communications and
  networking}, vol. 2018, no.~1, pp. 1--54, 2018.

\bibitem{PD_NOMA_cost}
M.~Baghani, S.~Parsaeefard, M.~Derakhshani, and W.~Saad, ``{Dynamic
  Non-Orthogonal Multiple Access and Orthogonal Multiple Access in 5G Wireless
  Networks},'' \emph{IEEE Trans. Commun.}, vol.~67, no. May, pp. 6360--6373,
  2019.

\bibitem{9152055}
C.~Deng, X.~Fang, X.~Han, X.~Wang, L.~Yan, R.~He, Y.~Long, and Y.~Guo, ``{IEEE
  802.11be {Wi-Fi} 7: New Challenges and Opportunities},'' \emph{IEEE Commun.
  Surv. \& Tutor.}, vol.~22, no.~4, pp. 2136--2166, Jul. 2020.

\bibitem{Mei1}
J.~Mei, X.~Wang, K.~Zheng \emph{et~al.}, ``{Intelligent Radio Access Network
  Slicing for Service Provisioning in {6G}: A Hierarchical Deep Reinforcement
  Learning Approach},'' \emph{IEEE Trans. Commun.}, pp. 1--1, Jun. 2021.

\bibitem{9217161}
A.~Ebrahim, A.~Celik, E.~Alsusa, and A.~M. Eltawil, ``{{NOMA/OMA} Mode
  Selection and Resource Allocation for Beyond 5G Networks},'' in \emph{Proc.
  IEEE Annual International Symposium on Personal, Indoor and Mobile Radio
  Communications (PIMRC'20)}, London, UK, Oct. 2020, pp. 1--6.

\bibitem{9148204}
I.~Khaled, C.~Langlais, A.~El~Falou, M.~Jezequel, and B.~ElHasssan, ``{Joint
  SDMA and Power-Domain NOMA System for Multi-User Mm-Wave Communications},''
  in \emph{International Wireless Communications and Mobile Computing (IWCMC)},
  July 2020, pp. 1112--1117.

\bibitem{MIMO-NOMA}
J.~{Wang}, Y.~{Li}, C.~{Ji}, Q.~{Sun}, S.~{Jin}, and T.~Q.~S. {Quek},
  ``{Location-Based MIMO-NOMA: Multiple Access Regions and Low-Complexity User
  Pairing},'' \emph{IEEE Trans. Commun.}, vol.~68, no.~4, pp. 2293--2307, Apr.
  2020.

\bibitem{8740921}
X.~Zhang, X.~Zhu, and H.~Zhu, ``{Joint User Clustering and Multi-Dimensional
  Resource Allocation in Downlink MIMO–NOMA Networks},'' \emph{IEEE Access},
  vol.~7, pp. 81\,783--81\,793, June 2019.

\bibitem{9461768}
Z.~Yang, M.~Chen, W.~Saad, and M.~Shikh-Bahaei, ``{Optimization of Rate
  Allocation and Power Control for Rate Splitting Multiple Access (RSMA)},''
  \emph{IEEE Trans. Commun.}, pp. 1--1, June 2021.

\bibitem{RSMA_JSAC}
L.~Zheng, Y.~Chencheng, C.~Ying, Y.~Sheng, and S.~Shlomo, ``{Rate Splitting for
  Multi-Antenna Downlink: Precoder Design and Practical Implementation},''
  \emph{IEEE J. Sel. Areas Commun.}, vol.~38, no.~8, pp. 1910--1924, Jun. 2020.

\bibitem{Liu_TWC}
Y.~{Liu}, X.~{Wang}, G.~{Boudreau}, A.~B. {Sediq}, and H.~{Abou-Zeid}, ``{A
  Multi-Dimensional Intelligent Multiple Access Technique for 5G Beyond and 6G
  Wireless Networks},'' \emph{IEEE Trans. Wireless Commun.}, vol.~20, no.~2,
  pp. 1308--1320, Feb. 2021.

\bibitem{Wudan}
W.~Han, J.~Mei, and X.~Wang, ``{User-Centric Multi-Dimensional Multiple Access
  in 6G Communications},'' in \emph{Proc. IEEE International Conference on
  Communications Workshops (ICC Workshops'21)}, Montreal, QC, Canada, 2021, pp.
  1--6.

\bibitem{8552437}
L.~{Sanguinetti}, A.~{Kammoun}, and M.~{Debbah}, ``{Theoretical Performance
  Limits of Massive MIMO With Uncorrelated Rician Fading Channels},''
  \emph{IEEE Trans. Commun.}, vol.~67, no.~3, pp. 1939--1955, Nov. 2019.

\bibitem{ding2020unveiling}
Z.~{Ding}, R.~{Schober}, and H.~V. {Poor}, ``{Unveiling the Importance of SIC
  in NOMA Systems: Part I -- State of the Art and Recent Findings},''
  \emph{IEEE Commun. Lett.}, vol.~24, no.~11, pp. 2373--2377, Jul. 2020.

\bibitem{JSAC_Yanan}
Y.~{Liu}, X.~{Wang}, J.~{Mei}, G.~{Boudreau}, H.~{Abou-Zeid}, and A.~B.
  {Sediq}, ``{Situation-Aware Resource Allocation for Multi-Dimensional
  Intelligent Multiple Access: A Proactive Deep Learning Framework},''
  \emph{IEEE J. Sel. Areas Commun.}, vol.~39, no.~1, pp. 116--130, Nov. 2021.

\bibitem{boehmer2020stable}
N.~Boehmer and E.~Elkind, ``{Stable Roommate Problem with Diversity
  Preferences},'' \emph{arXiv preprint arXiv:2004.14640}, Apr. 2020.

\bibitem{8675293}
Y.~Li, Y.~Jiang, W.~Wu, J.~Jiang, and H.~Fan, ``{Room Allocation With Capacity
  Diversity and Budget Constraints},'' \emph{IEEE Access}, vol.~7, pp.
  42\,968--42\,986, Mar. 2019.

\bibitem{2016transfers}
E.~Lazarova, P.~Borm, and A.~Est{\'e}vez-Fern{\'a}ndez, ``{Transfers and
  Exchange-Stability in Two-Sided Matching Problems},'' \emph{Theory and
  Decision}, vol.~81, no.~1, pp. 53--71, June 2016.

\bibitem{SCA}
J.~Papandriopoulos and J.~S. Evans, ``{Low-Complexity Distributed Algorithms
  for Spectrum Balancing in Multi-User {DSL} Networks},'' in \emph{Proc. IEEE
  International Conference on Communications (ICC'06)}, vol.~7, Istanbul,
  Turkey, 2006, pp. 3270--3275.

\bibitem{Convex}
S.~Boyd and L.~Vandenberghe, \emph{Convex Optimization}.\hskip 1em plus 0.5em
  minus 0.4em\relax Cambridge, U.K.: Cambridge University Press, 2004.

\bibitem{JSAC_1995}
R.~Yates, ``{A Framework for Uplink Power Control in Cellular Radio Systems},''
  \emph{IEEE J. Sel. Areas Commun.}, vol.~13, no.~7, pp. 1341--1347, 1995.

\end{thebibliography}
%
%
\appendix
\subsection{Proof of \textbf{\textit{Proposition 1}}}
\begin{IEEEproof}
 {To prove the convergence of the rotation process in \textbf{Algorithm 1}, we denote the coalition formation result during the $\lambda$-th iteration as $\Pi_{\lambda}$ without loss of generality. Provided that $\xi^{*}_{s}$ is a valid and optimal rotation sequence under the same selected \begin{small}$K_{r}$\end{small} in $\left(\lambda+1\right)$-th iteration, which is equivalently to \begin{small}$\mathcal{G}^{\rm{cost}}_{\Pi_{\mathcal{K}_r,\xi^{*}_s}}\leq\mathcal{G}^{\rm{cost}}_{\Pi_{\lambda}}$\end{small}, then during the next round of iteration, the coalition formation changes from \begin{small}$\Pi_{\lambda}$\end{small} to \begin{small}$\Pi_{\lambda+1}$\end{small} and we have \begin{small}$\Pi_{\lambda+1}=\Pi_{\mathcal{K}_r,\xi^{*}_s}$\end{small}. Therefore, the total resource sharing conflicts is guaranteed to decrease after each rotation operation, and it will ultimately terminate with a rotation sequence $\xi^{*}_{s}$ as a convergence point. In other words, no any other rotation operation can further reduce the utility value defined in $\mathcal{P}_1$. 
 \par
 The concept of exchange stability under \textit{rotation sequence} context presumes that the only possibility for deviation in current coalition formation is that there still exists some switches (i.e., UEs exchange their matched coalitions, or coalitions swap their matched UEs) provided that such operations meet the coalition criteria and can create a better-off situation for any affected parties (UEs or coalitions). In our case, such exchange can happen only if it satisfies constraint C1 as well as C2, and meantime, provides an improvement for $\mathcal{P}_{1}$. Otherwise, the coalition formation result is acknowledged to achieve the exchange stability, which means no further swap (i.e., \begin{small}$S=2$\end{small}) or even more complicated exchanges (i.e., \textit{rotation sequence} operations when \begin{small}$S>2$\end{small}) can be performed to earn a lower value defined in $\mathcal{P}_{1}$. Given the proved convergence of \textbf{Algorithm 1}, we can say that the final coalition formation is deemed to be exchange-stable.  
}
\end{IEEEproof}
\end{document}